# Chemical and interface confinement effects in promoting plastic co-deformation in high-strength nano-scale eutectics

Arkajit Ghosh[#], Amit Misra[#]

Department of Materials Science and Engineering, University of Michigan – Ann Arbor, MI 48109, USA

[#] Corresponding authors: arkajitg@umich.edu, amitmis@umich.edu

## *Abstract*

Eutectics offer a route to overcome plastic incompatibility in disparate-phase heterostructures by coupling microstructural refinement with phase-specific chemical and crystallographic hierarchy. Here, we investigate laser-rapid-solidified Al–(Si,Ge) eutectic composites designed along the univariant ternary eutectic path, solidifying with an Al-rich face-centered cubic (*fcc*) matrix and (Si,Ge)-rich diamond-cubic (*dc*) fibers as constituent eutectic phases with faceted interfaces. The microstructure was hierarchical in nature with finer structures within eutectic phases: nanoscale (Si,Ge) clusters in the Al phase and growth twins within the (Si,Ge) fibers with some Al retention. Although increasing Ge content coarsens the eutectic spacing, the yield strength is slightly enhanced compared to the relatively finer Al-Si, and tensile ductility of Al–(Si,Ge) is higher than that of Al–Si. *In situ* SEM micromechanical testing combined with *post-mortem* STEM and TEM showed that in the Al-rich phase, Ge segregated to the deformation-induced sub-grain boundaries that confined glide dislocations. Simultaneously, the (Si,Ge)-rich fibers remain crack-resistant at large plastic strain and exhibit deformation-induced planar faults, consistent with localized partial-dislocation activity at highly stressed interfaces and twin boundaries. These coupled mechanisms due to interface confinement and hard phase chemistry enable co-deformation of the metallic and covalent phases, providing a pathway for designing high-tensile-strength and ductile hierarchical eutectics beyond conventional length-scale-controlled strengthening.

### Graphical abstract

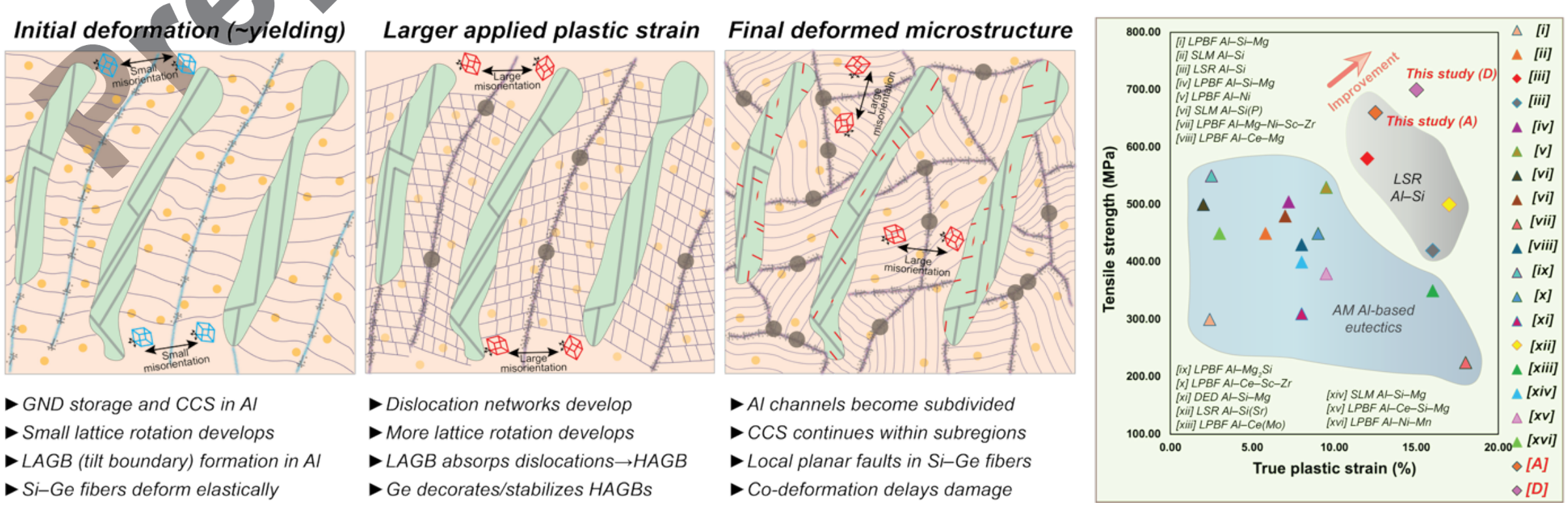

## 1. Introduction

Eutectic composites are attractive structural materials because coupled eutectic growth naturally produces periodic soft/hard phase microstructures with high interface densities [1, 2, 3, 4]. In regular eutectics, the characteristic lamellar or rod spacing is selected by the coupled-growth condition and scales with the solidification undercooling and growth rate, so rapid solidification can drive the microstructure into the nanometer regime [5, 6]. When the characteristic phase spacing is reduced to the nanoscale, the interfaces impose geometric confinement on dislocation motion. The soft phase can no longer deform as an unconstrained continuum; instead, its plasticity becomes governed by interface-affected confined dislocation motion, and strong dislocation storage at neighboring phase boundaries [7, 8, 9]. This idea was formalized for nanoscale layered composites that showed flow strength rises as the available slip distance decreases and as interfaces increasingly control dislocation transmission and storage [10, 11]. Rapid-solidified Al–Si has been used as a model system to explore micromechanical response of periodic microstructure with disparate soft/hard phases: laser rapid solidification produces ultrafine, interconnected fibrous Al–Si eutectics with high yield strength, substantial strain hardening and tensile ductility [12, 13]. Yet refinement alone does not solve the central problem of plastic incompatibility, as in Al–Si eutectics, plastic strain is mainly carried by the Al channels, while the covalent Si phase remains difficult to plastically accommodate at room temperature [14]. Atomistic modeling has shown that Al–Si interfaces possess low shear resistance and low defect-formation/migration barriers, so dislocations approaching the interface can cross-slip (screw segments) onto it or climb (edge segments) along it and eventually reflect back into the Al phase rather than transfer across the interface into Si [15]. This relieves local stresses in Al but weakens the load-bearing contribution of the Si reinforcement. Experimentally, continued straining of laser-rapid-solidified Al–Si leads to closely spaced dislocation arrays in Al, segmentation/fracture of Si fibers, and recovery in Al, making brittle failure of the hard phase the key obstacle to true plastic co-deformation [16, 17].

Several strategies have been explored to overcome the plastic incompatibility of Al–Si eutectics by modifying the morphology, chemistry, and defect structure of the covalent Si phase. Impurity modification using elements such as Sr, Ce, and other rare-earth additions can transform coarse faceted Si into finer fibrous or branched morphologies and may introduce growth twins and stacking faults in the Si phase through a mechanism called impurity induced twinning often mediated by solute/impurity clusters [18, 19, 20]. Other alloying approaches, such as Cu addition, can improve the strength of Al–Si-based eutectics, but they commonly introduce additional brittle intermetallic/eutectic constituents such as $Al_2Cu$ and produce an Al–$Al_2Cu$–Si multiphase microstructure rather than preserving a two-phase Al/Si-like eutectic topology [21, 22]. These prior studies show that chemical modification can strongly affect the morphology and deformation response of Al-based eutectics, but most strategies either modify the Si phase morphology from flake to fiber or introduce additional brittle phases. Therefore, a fundamental question remains: can eutectic microstructures in ternary alloys exhibit synergistic effects due to interface confinement and alloy chemistry to modify the deformation mechanisms and enhance plasticity in the hard phase?

In this work, we address these scientific issues by designing a series of univariant Al–Si–Ge nano-scale eutectics through laser rapid solidification (LRS) and systematically varying composition along the univariant ternary eutectic tie-line in the ternary Al–Si–Ge phase diagram. Ge is attractive for this purpose for several reasons. First, Si and Ge form an isomorphous diamond-cubic (*dc*) solid solution, and the Al–Si–Ge system contains a univariant Al – (Si,Ge) eutectic path, allowing composition to be varied without leaving the desired two-phase topology [23]. Second, this avoids the rapid-solidification behavior of binary Al–Ge, where the stable Al + Ge eutectic would be replaced by brittle metastable Al–Ge intermetallic eutectics at high growth velocity [24, 25]. Third, Ge incorporation into Si may reduce the barrier for defect motion in the covalent phase because Ge has a lower Peierls stress than Si [26]. The deformation mechanisms in these unique hierarchical structures are experimentally characterized and discussed.

## 2. *Experimental methods*

### *(i) Materials and processing*

The eutectic compositions were designed based on the univariant ternary eutectic tie-line in the Al–(Si,Ge) system, shown in Figure 1a, along which a two-phase Al and (Si,Ge) eutectic morphology is expected to be retained. Under rapid, non-equilibrium solidification, however, the effective eutectic composition can shift toward the hypereutectic side due to displacement of the liquidus lines toward the corresponding $T_0$ (temperature at which liquid and solid phases of the same composition have equal Gibbs free energy due to complete solute trapping conditions) lines [5, 27]. Therefore, hypereutectic nominal compositions were selected, as illustrated in Figure 1b. This strategy follows prior work on laser-rapid-solidified Al–Si eutectics, where increasing the nominal composition from the equilibrium binary eutectic composition, Al–12.7 wt.% Si, to Al–20 wt.% Si produced a fully eutectic microstructure after rapid solidification [28]. Accordingly, a modified hypereutectic composition path was constructed by drawing a line parallel to the equilibrium Al–(Si,Ge) eutectic tie-line, shown as the red line in Figure 1a. This modified tie-line intersects the Al–Si binary at Al–20 wt.% Si and the Al–Ge binary at approximately Al–65 wt.% Ge, thereby defining the nominal composition range investigated in this study.

Four eutectic composites, denoted A–D, were prepared by arc melting according to this modified hypereutectic tie-line (microstructures have been reported in Supplementary Figure S1). Their nominal compositions are summarized in Table 1. Laser surface remelting was then performed on the arc-melted samples using a PANDA$^{TM}$ machine, as schematically illustrated in Figure 1c, to generate rapidly solidified melt pools containing ultrafine eutectic microstructures (for the actual melt pool images, refer to Supplementary Figure S2). The laser power, scanning speed, and spot diameter were 200 W, 100 mm/s, and 75 μm, respectively. For compositions with Ge contents higher than that of D, additional metastable intermetallic phases were observed after laser remelting. Therefore, the present study was restricted to compositions A–D to preserve the desired two-phase Al + (Si,Ge) eutectic heterostructure.

Table 1: Prepared Al–(Si,Ge) compositions

| Composition | Al (wt.%) | Si (wt.%) | Ge (wt.%) |
|---|---|---|---|
| A | 80.0 | 20.0 | 0.0 |
| B | 74.4 | 17.5 | 8.1 |
| C | 68.7 | 15.0 | 16.3 |
| D | 57.5 | 10.0 | 32.5 |

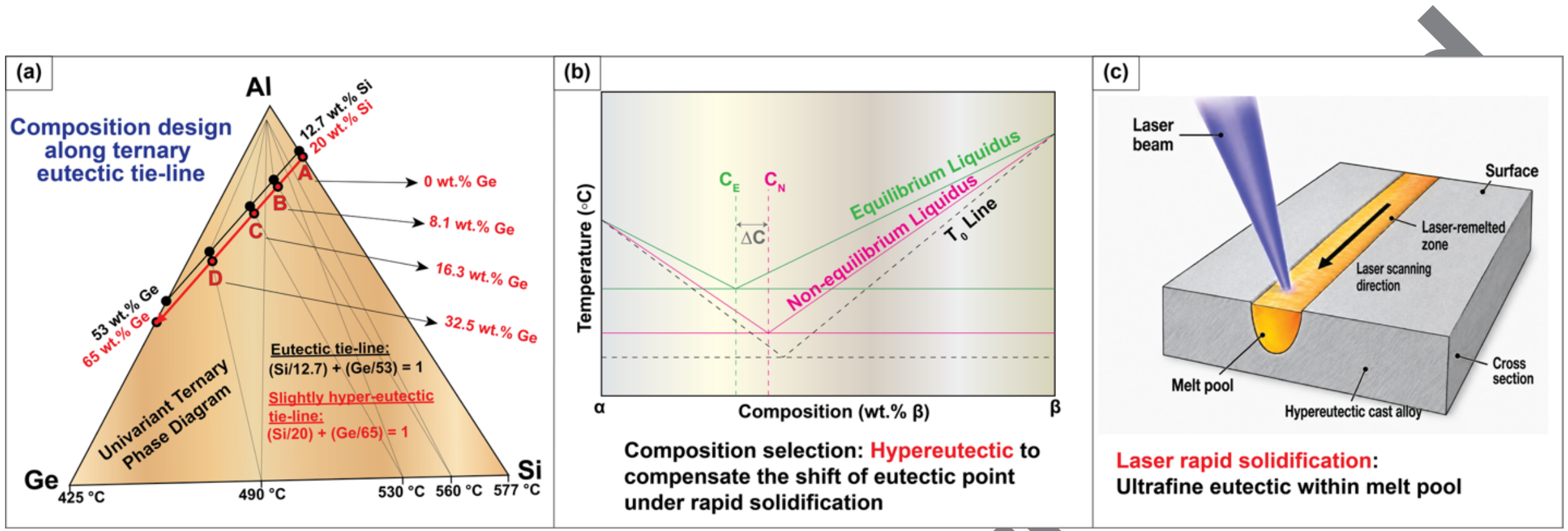

**Figure 1. Composition design and laser surface remelting of Al–(Si,Ge) eutectic composites.** (a) Al–(Si,Ge) ternary phase diagram showing the equilibrium Al + (Si,Ge) univariant eutectic tie-line and the modified hypereutectic composition path used for eutectics A–D. (b) Schematic showing the rapid-solidification-induced shift (*ΔC*) in effective eutectic composition from $C_E$ to $C_N$, motivating the selection of hypereutectic nominal compositions. (c) Laser surface remelting schematic showing formation of an ultrafine eutectic microstructure within the melt pool.

### *(ii) Microstructure characterization*

Microstructures were examined by scanning electron microscopy (SEM) using a TFS Helios 650 NanoLab in BSE mode at 5 kV and 0.2 nA. Site-specific TEM specimens were prepared by Xe plasma focused ion beam (FIB) lift-out and thinned to ~30 nm. Atomic-resolution scanning transmission electron microscopy (STEM) was performed on an aberration-corrected TFS Spectra operated at 300 kV and 75 pA screen current. STEM-EDX mapping was conducted with a Dual-X detector at 100 pA to improve elemental signal. STEM tomography experiments were conducted using a Fischione high-tilt high-visibility tomography holder and tilting the goniometer from −70° to +70° at an interval of 2° followed by three-dimensional reconstruction using Gatan GMS 3 software [29].

### *(iii) Micromechanical testing*

*In situ* micropillar compression and dog-bone tensile testing were performed inside a TESCAN MIRA3 SEM using a Hysitron PI 89 Picoindenter. Cylindrical micropillars with nominal dimensions of ~8 μm diameter and ~18 μm height, and dog-bone tensile specimens with gauge dimensions of ~10.5 μm × 3.5 μm × 3.5 μm, were fabricated using Xe plasma FIB milling in a TFS Hydra 5 PFIB-SEM following optimized geometries reported previously [30]. Compression tests were carried out using a Bruker 20 μm flat-punch, 60 μm conical diamond probe mounted on a high-load transducer. Tensile tests were performed using a FIB-milled diamond gripper prepared from a Berkovich indenter, with a jaw opening of ~10 μm. All tests were conducted under displacement control at a nominal strain rate of 0.5% /s up to a maximum nominal strain of 30%. Four compression and two tensile tests were performed for each condition to assess reproducibility. Representative true stress–strain curves are shown in the main text, while engineering stress–strain curves for all individual tests are provided in the Supplementary Information. Engineering stress and strain were calculated from the measured load–displacement data using the pre-loading specimen dimensions determined by SEM, and the corresponding true stress–strain curves were subsequently derived from these data.

### *(iv)* ***Theoretical calculations***

First-principles calculations were performed using density functional theory (DFT) as implemented in Quantum ESPRESSO [31], using the PBE generalized-gradient approximation [32], PSLibrary pseudopotentials [33], and Monkhorst–Pack k-point sampling [34]. Three complementary calculations were performed. First, Ge segregation in Al was evaluated using the experimentally identified $\sum 19a\ \{331\}\langle 110\rangle$ symmetric tilt boundary in a 74-atom periodic bicrystal containing two equivalent boundaries separated by $\sim 35.2$ Å. Five Ge substitution sites near the boundary (GB1–GB5) were compared with a bulk-like site $\sim 17.61$ Å away using 50/300 Ry wavefunction/charge-density cutoffs and a $4 \times 2 \times 1$ k-point mesh. The most favorable configuration and bulk reference were fully relaxed to a force criterion of $2 \times 10^{-3}$ Ry/Bohr. Second, the effect of retained Al on Si shear resistance was examined using 64-atom Si and $Si_{63}Al$ supercells with 50/400 Ry cutoffs and a $2 \times 2 \times 2$ k-point mesh. Homogeneous shear was imposed on the $(111)[1\bar{1}0]$ system at $\gamma = 0$, 0.05, 0.10, and 0.15 using fixed-ion calculations. Third, generalized stacking-fault-energy (GSFE) calculations were performed for pure Si and a representative Si–50Ge configuration using 96-atom supercells. The calculations employed plane-wave kinetic-energy/charge-density cutoffs of 45/300 Ry. The upper portion of the crystal was progressively displaced relative to the lower portion along the glide-set $(111)\ {}^{a}/_{6}\langle 112\rangle$ Shockley-partial pathway at normalized displacements $\frac{u}{b_p} = 0$, 0.25, 0.5, 0.75, and 1. Atomic positions were relaxed at each imposed displacement while maintaining the prescribed relative displacement across the fault plane, with ionic relaxation continued until the total force was below approximately $5 \times 10^{-3}$ Ry/Bohr.

## 3. *Results*

### *(i)* ***As-processed microstructure: Length-scale, morphology, and hierarchy***

Figure 2(a–d) shows the as-processed microstructures of the laser-remelted eutectic composites. Composition A corresponds to the binary Al–Si reference eutectic, while compositions B–D are univariant ternary Al–(Si,Ge) alloys with progressively increasing Ge

content. SEM images acquired from the rapidly solidified melt pools reveal that all four compositions develop ultrafine eutectic microstructures. This indicates that the modified hypereutectic composition path successfully preserves the desired two-phase Al + Si or Al + (Si,Ge) eutectic topology after laser rapid solidification. Although the two-phase eutectic morphology is retained from the binary Al–Si reference eutectic to the highest-Ge eutectic, the characteristic eutectic length scale and volume fraction of second/fiber phase change systematically with Ge addition. The SEM images show progressive coarsening from eutectic A to eutectic D. Quantitative measurements [Figure 2(e–f)] confirm this trend: the average fiber diameter increases from $32 \pm 4$ nm in eutectic A to $51 \pm 6$ nm in eutectic D, while the average inter-fiber spacing increases from $52 \pm 7$ nm to $98 \pm 10$ nm. In parallel, the estimated fiber-/second-phase volume fraction, ($V_f$) increases from ~15% in eutectic A to ~19%, ~24%, and ~33% in eutectics B, C, and D, respectively. This composition-dependent coarsening occurs even though all samples were processed using the same laser power, scan speed, and beam diameter, indicating that the selected eutectic length scale is controlled not only by the imposed processing condition but also by changing thermodynamics and kinetics of the ternary liquid/solid interface induced by Ge addition.

The coarsening with increasing Ge content can be understood from eutectic growth selection in a multicomponent system. In a regular eutectic, the selected spacing reflects a balance between solutal undercooling and curvature undercooling [35]. Finer spacings reduce solute diffusion distances but increase interfacial curvature, whereas coarser spacings reduce curvature undercooling but require longer-range solute redistribution [36]. Along the Al–(Si,Ge) eutectic path, increasing Ge changes the liquidus slopes, eutectic temperature, solute partitioning, and diffusional field ahead of the solidification front. Unlike binary Al–Si, where coupled growth mainly requires redistribution of Si, ternary Al–(Si,Ge) solidification requires coupled redistribution of both Si and Ge between the Al-rich *fcc* and the (Si,Ge)-rich *dc* phases [23, 37]. The increasing Ge/Si ratio therefore changes the minimum-undercooling spacing selected by the coupled-growth front, leading to a coarser eutectic length scale at higher Ge contents. This coarsening is also consistent with the lower melting/eutectic temperatures associated with Ge-rich compositions [Figure 1a] and the increased chemical complexity of the solutal boundary layer during rapid solidification. Under nominally identical laser parameters, the local thermal gradient and interface velocity may be similar, but the effective growth undercooling, freezing range, and solute trapping tendency can vary with alloy composition [38]. As Ge content increases, the coupled eutectic front must accommodate a larger solute redistribution requirement while maintaining the Al + (Si,Ge) two-phase topology. The observed increase in both fiber diameter and inter-fiber spacing from A to D therefore reflects a composition-driven shift in the stable coupled-growth length scale rather than a breakdown of eutectic growth.

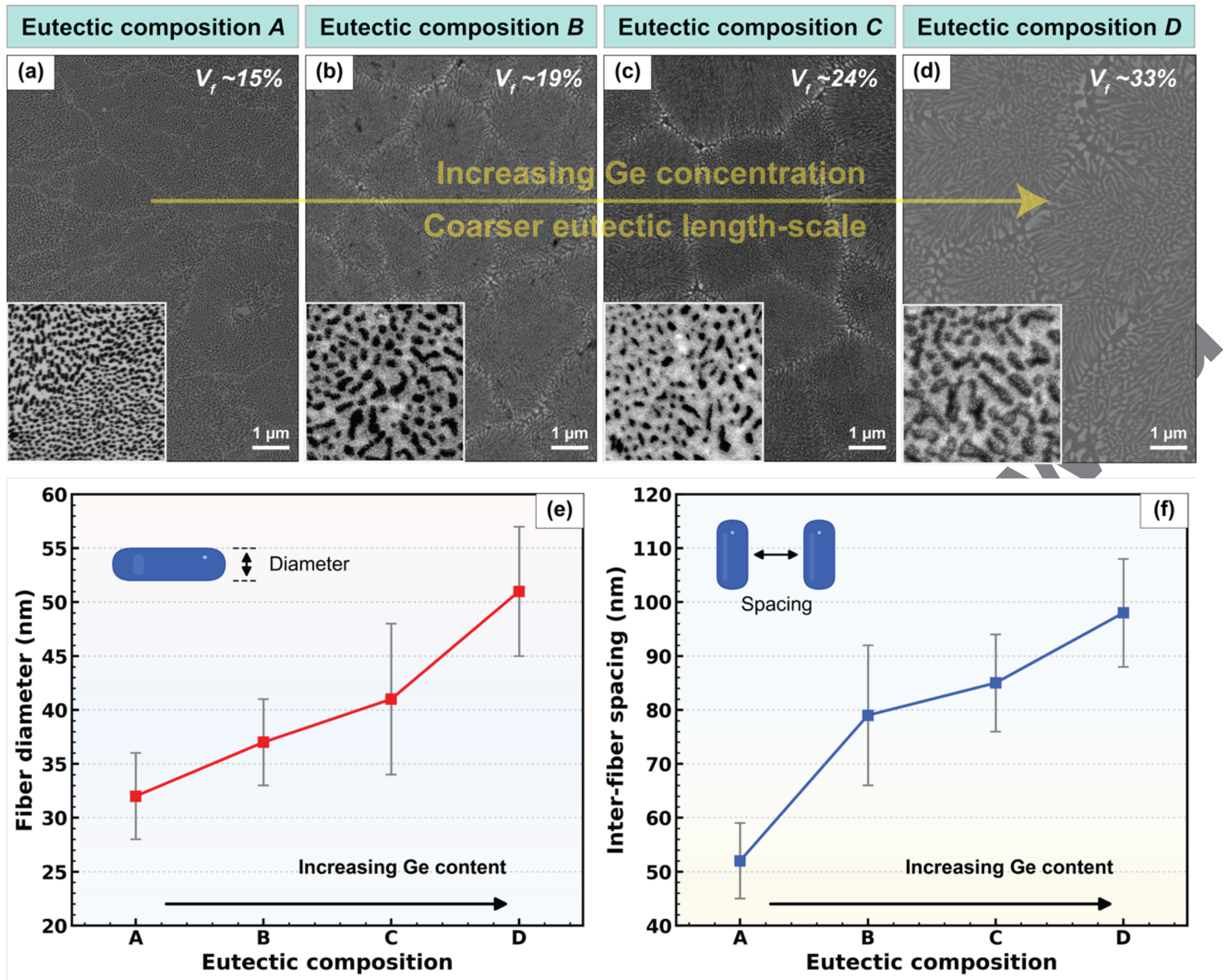

**Figure 2: As-processed eutectic microstructures and composition-dependent length-scale evolution after laser rapid solidification.** (a–d) SEM images of laser-remelted eutectic composites A–D with higher magnification (8x) secondary electron micrographs as insets. Composition A is the binary Al–Si reference eutectic, while B–D are univariant ternary Al–(Si,Ge) eutectics with progressively increasing Ge content (Table 1). The two-phase eutectic morphology is retained from A to D, but the characteristic length scale coarsens systematically with increasing Ge content. (e) Average fiber diameter and (f) average inter-fiber spacing as a function of eutectic composition. The fiber diameter increases from 32 ± 4 nm in A to 51 ± 6 nm in D, while the inter-fiber spacing increases from 52 ± 7 nm to 98 ± 10 nm, indicating a composition-driven shift in eutectic growth selection under identical laser-processing conditions.

Figure 3 confirms the two-phase nature of the eutectic microstructure at higher spatial and chemical resolution. STEM-HAADF imaging and corresponding STEM-EDX elemental maps in Figure 3a (presented from composition D only – the similar observation obtained from other compositions have been reported in Supplementary Figure S3) show complementary partitioning of Al, Si, and Ge. The matrix is enriched in Al, while the fibrous phase is enriched in both Si and Ge. The STEM-EDX line profile across an individual fiber shows a sharp decrease in Al signal and a simultaneous increase in Si and Ge signals within the fiber, confirming chemical partitioning across the Al/(Si,Ge) interface. These results support the initial observation from SEM that the microstructure remains a two-phase Al/(Si,Ge) eutectic rather than developing separate Si-rich, Ge-rich, or intermetallic phases.

A reconstructed STEM tomography volume from composition D [Figure 3b] further demonstrates that the (Si,Ge)-rich phase forms a three-dimensional rod/fiber network embedded within the Al-rich matrix. This analysis is important because two-dimensional micrographs can make the eutectic morphology appear ambiguous, particularly when rods, fibers, or interconnected branches are sectioned at different orientations. The tomographic reconstruction shows that the (Si,Ge)-rich phase is not an assembly of isolated particles, but a connected eutectic phase distributed through the Al matrix. This morphology is similar to the fibrous Si network observed in rapidly solidified Al–Si eutectics [39], indicating that the Al–(Si,Ge) eutectics remain structurally comparable to the binary Al–Si reference despite the increased Ge content, larger eutectic length scale, and higher (Si,Ge)-rich phase fraction.

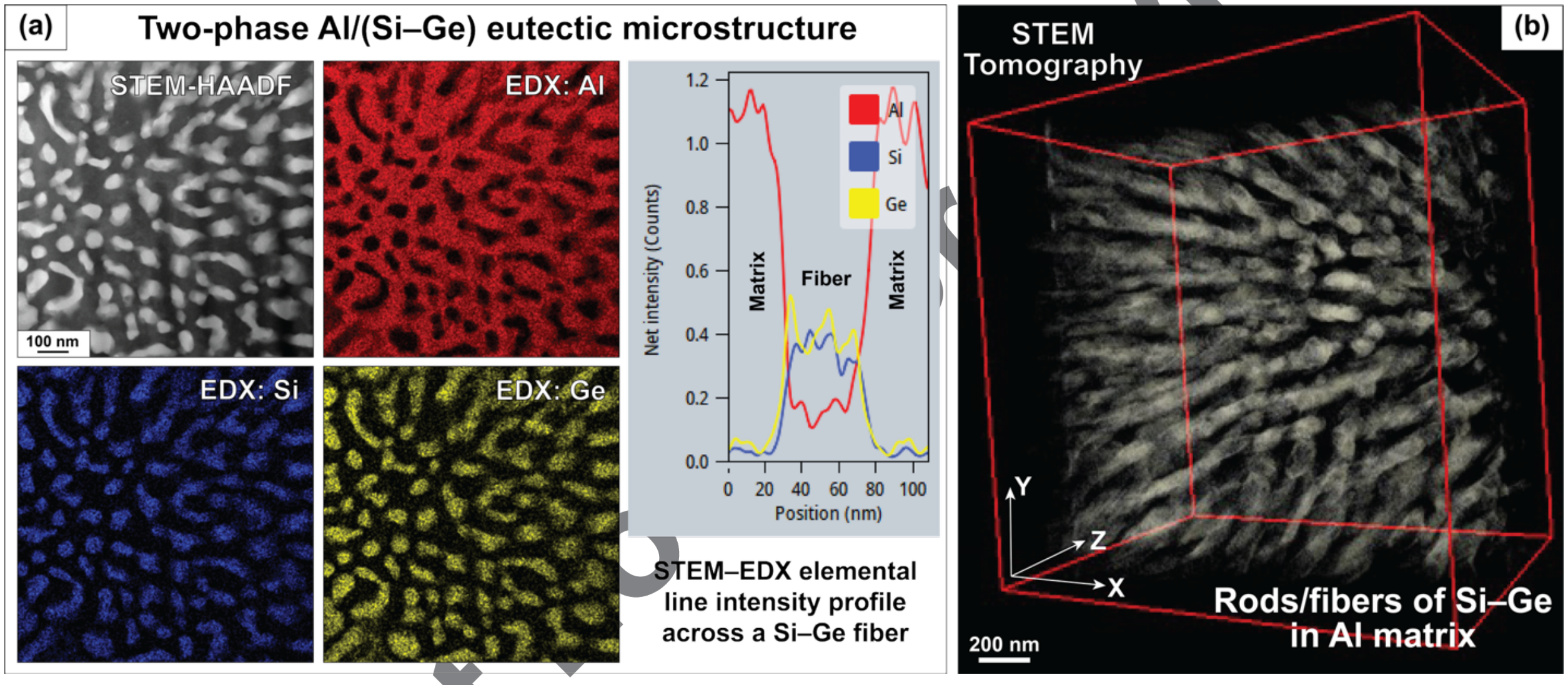

**Figure 3: Two-phase Al/(Si,Ge) eutectic morphology and chemical partitioning.** (a) STEM-HAADF image, STEM-EDX elemental maps, and representative line profile from composition D showing an Al-rich matrix and (Si,Ge)-rich fibers. The complementary Al, Si, and Ge distributions confirm two-phase Al/(Si,Ge) partitioning without separate Si-rich, Ge-rich, or intermetallic phases. (b) STEM tomography reconstruction from composition D showing a connected three-dimensional rod/fiber network of the (Si,Ge)-rich phase embedded within the Al matrix, similar to the reference Al–Si eutectic.

The persistence of the rod/fiber morphology in the highest-Ge alloy is particularly notable because the estimated (Si,Ge)-rich phase fraction is close to the geometric threshold at which a rod-to-lamella transition may be expected in an ideal eutectic. In a simplified interfacial-energy argument, the stability of rod-like and lamellar morphologies depends on the volume fraction of the second phase, with rod-like eutectics favored below a critical fraction of approximately ($1/\pi \approx$ 0.32), whereas larger second-phase fractions increasingly favor lamellar arrangements [2, 5, 40]. Therefore, the increased (Si,Ge)-rich phase fraction in composition D places the alloy above the regime where lamellar growth could be thermodynamically competitive. However, this volume-fraction criterion is only a first-order guideline because it assumes isotropic interfacial energies, uniform spacing, and near-equilibrium morphology selection. Under laser rapid solidification, morphology selection is strongly affected by kinetic factors, including high interface velocity, restricted long-range solute diffusion, solute trapping, and anisotropic interface attachment of the faceted *dc* (Si,Ge) phase [41, 42]. These kinetic effects can preserve or stabilize a fibrous growth front even when the phase fraction approaches the range where lamellae would be favored in an ideal isotropic eutectic [43]. Thus, the observed rod/fiber morphology in composition D reflects a competition between increasing second-phase fraction, which tends to favor lamellar growth, and rapid-solidification/faceted-growth kinetics, which favor branching and rod-like (Si,Ge) growth within the non-faceted Al matrix. Ge addition, therefore, changes the eutectic length scale and phase chemistry without producing a rod-to-lamella transition, enabling direct comparison between Al–Si and Al–(Si,Ge).

After establishing the two-phase Al/(Si,Ge) eutectic morphology and its composition-dependent length-scale evolution, we next examine the internal chemical and crystallographic hierarchy within each constituent phase. In addition to the primary eutectic partitioning between the Al-rich matrix and the Si- or (Si,Ge)-rich fibers, laser rapid solidification produces nanoscale solute heterogeneity within the Al phase and defect structures within the *dc* phase. These features are important because they introduce additional length scales and potential barriers or nucleation sites for defects, beyond the eutectic spacing alone, when subjected to deformation.

Figure 4 shows the development of nanoscale chemical hierarchy within the Al-rich phase as the Ge content increases. Figure 4a shows an atomic-resolution STEM-HAADF image from eutectic D, where a high density of nanoscale clusters is observed within the Al-rich *fcc* matrix. The clusters are coherent with the surrounding Al lattice, as indicated by the continuity of the lattice fringes across the clustered regions and by the FFT indexed to the *fcc* Al matrix. STEM-EDX maps acquired from the marked region in Figure 4a show local enrichment of both Si and Ge within the clusters, confirming that these features correspond to (Si,Ge) solute clusters rather than structural contrast alone.

Figure 4(b–c) shows the composition dependence of this clustering behavior. In eutectic B, Figure 4b shows a relatively uniform Al-rich matrix in both STEM-HAADF imaging and EDX elemental maps, with no clear evidence of nanoscale (Si,Ge) clustering. In contrast, eutectic C in Figure 4c exhibits localized regions of correlated Si and Ge enrichment within the Al-rich phase, indicating the onset of (Si,Ge) clustering. Therefore, the clustering tendency increases systematically with Ge content: no clear clustering is detected in eutectic B, clustering begins to appear in eutectic C, and a high density of coherent clusters is present in eutectic D.

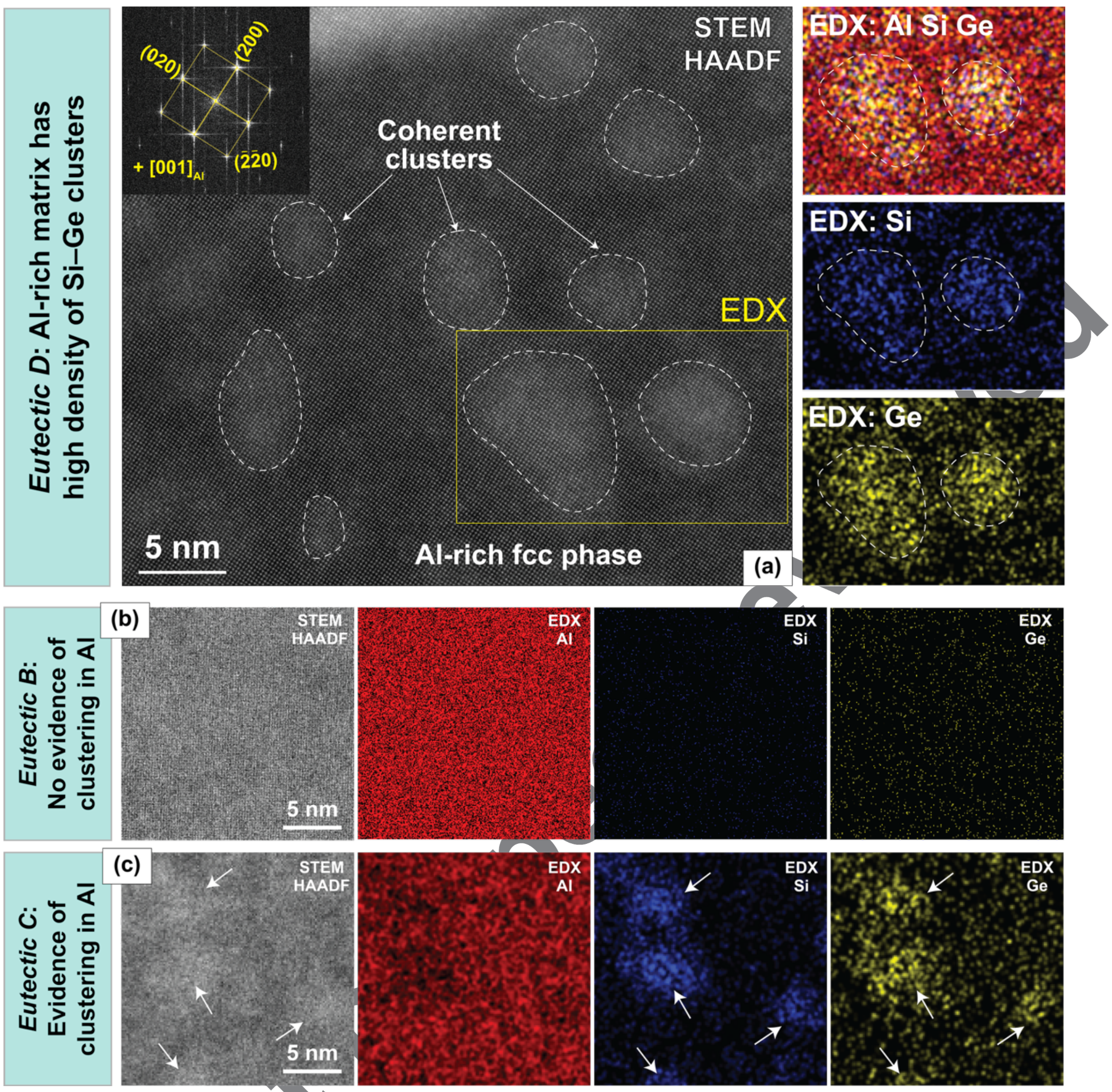

**Figure 4. Nanoscale (Si,Ge)-rich clustering in the Al-rich phase.** (a) Atomic-resolution STEM-HAADF image from eutectic D showing coherent nanoscale clusters embedded within the Al-rich *fcc* matrix. The inset FFT confirms the *fcc* Al structure, and the corresponding STEM-EDX maps from the marked region show local Si and Ge enrichment within the clusters. (b) STEM-HAADF image and EDX maps from eutectic B showing a relatively uniform Al-rich matrix with no clearly resolved (Si,Ge)-rich clusters. (c) STEM-HAADF image and EDX maps from eutectic C showing the onset of correlated Si and Ge enrichment within the Al-rich phase. Therefore, the clustering tendency increases with Ge content, becoming most pronounced in eutectic D.

The formation of these clusters can be understood as a consequence of rapid-solidification-induced supersaturation followed by short-range solute decomposition within the Al-rich phase. Under equilibrium conditions, the solubility of both Si (~1.5 wt.%) and Ge (~5.5 wt.%) in *fcc* Al is limited at respective eutectic temperatures, and these solutes are expected to partition primarily to the *dc* eutectic phase. During laser rapid solidification, however, the high interface velocity restricts long-range solute redistribution and allows partial solute trapping within the growing Al-rich phase [44, 45]. The extent of this trapped supersaturation increases with Ge addition because the ternary liquid must redistribute both Si and Ge during coupled eutectic growth [23]. At low Ge content, as in eutectic B, the retained Si and Ge concentration in Al is likely below the threshold required for detectable clustering, and the Al-rich matrix remains chemically uniform at the resolution of STEM-EDX. In eutectic C, the increased Ge content raises the local Si + Ge supersaturation sufficiently to initiate correlated (Si,Ge) clustering, while in eutectic D the higher Ge/Si ratio produces a stronger chemical driving force for nanoscale solute decomposition, resulting in a high density of clusters. The correlated enrichment of Si and Ge indicates that clustering is not caused by independent precipitation of each solute, but by coupled chemical association between Si and Ge within the supersaturated Al matrix. Because Si and Ge are mutually soluble in the diamond-cubic phase, local co-enrichment provides a favorable pathway to reduce the free energy of the supersaturated Al solid solution without forming a fully developed second phase during rapid quenching [46]. The coherent nature of the clusters further suggests that they remain structurally constrained by the surrounding *fcc* Al lattice, minimizing interfacial energy and suppressing coarsening [47]. Therefore, increasing Ge content introduces a composition-dependent internal length scale within the Al-rich phase, distinct from the eutectic spacing, that becomes most pronounced in eutectics C and D.

The internal hierarchy of the Si- or (Si,Ge) fiber phase is shown in Figure 5. Figure 5(a1–a2) present representative STEM-HAADF images from eutectic D, where $\{111\}\langle 11\bar{2}\rangle$ growth twins extend through the (Si,Ge) fibers. Similar twin-containing fibers are observed across the other eutectics, as shown in the Supplementary Figure S5. The high density of twins can be rationalized by the twin-plane re-entrant edge (TPRE) growth mechanism of faceted diamond-cubic Si and Ge phases, where twin boundaries provide favorable re-entrant sites for atom attachment, lateral propagation, and branching during eutectic growth [48]. Compared with binary Al–Si, where growth-induced stacking faults are commonly observed in Si, the Ge-containing fibers show predominantly twin-like planar defects, which suggests that Ge addition modifies the effective fault/twin formation energetics and faceted growth kinetics of the *dc* phase during rapid ternary eutectic solidification. Figure 5b shows that the Al/(Si,Ge)interface is faceted and contains atomic-scale terraces and ledges. This interfacial morphology is consistent with TPRE-mediated growth and the intersection of twin boundaries with the eutectic interface. The FFT from the (Si,Ge)-rich fiber confirms the *dc* structure, whereas the adjacent Al matrix is not aligned along a nearby low-index zone axis under the same imaging condition. Additional attempts to orient both phases along different zone axes within the same eutectic colony did not reveal a reproducible crystallographic orientation relationship (OR) between the Al matrix and the (Si,Ge)-rich fibers. Such an absence of a unique colony-scale OR is reasonable for rapidly solidified fibrous eutectics, where local faceting, repeated twinning, branching, and high growth velocity can produce spatially varying interfacial configurations [49]. This interpretation is in good agreement with prior reports showing random or variable Al/group-IV phase relationships in Al–Ge and rapidly solidified Al–Si eutectics, although specific local ORs can occur in some Al–Si eutectic structures [50, 51]. The interface appears to be governed locally by faceting along the $\{111\}_{Si}$ planes, ledge formation, and crystallographic anisotropy of the *dc* phase rather than by a unique Al/(Si,Ge)OR.

In addition to growth twins and faceted interfaces, STEM-EDX analysis of composition D shows that the (Si,Ge)-rich fibers contain a measurable Al signal [Figure 5(c1–c5)]. The Al distribution within the fibers appears spatially diffuse within the resolution of the STEM-EDX maps, with no obvious evidence of clustering of Al. This feature is better described as Al retention, or rapid-solidification-induced solute trapping, within the *dc* (Si,Ge)-rich phase rather than as a separate Al-rich precipitate or cluster population [52, 53], also reported in prior work on pure and Sr modified Al–Si processed by laser rapid solidification [20]. Under equilibrium conditions, Al has limited solubility in *dc* Si and (Si,Ge); however, the high interface velocity during eutectic growth can restrict long-range redistribution of Al away from the advancing *dc* solidification front. As a result, a finite amount of Al is retained within the growing Si- or (Si,Ge)-rich fibers, particularly near growth defects, ledges, and twin boundaries. The faceted interfaces and dense growth twins shown in Figure 5(a1, a2, b) may therefore provide additional local sites for Al incorporation during growth, contributing to the chemical and defect-level hierarchy within the fiber phase.

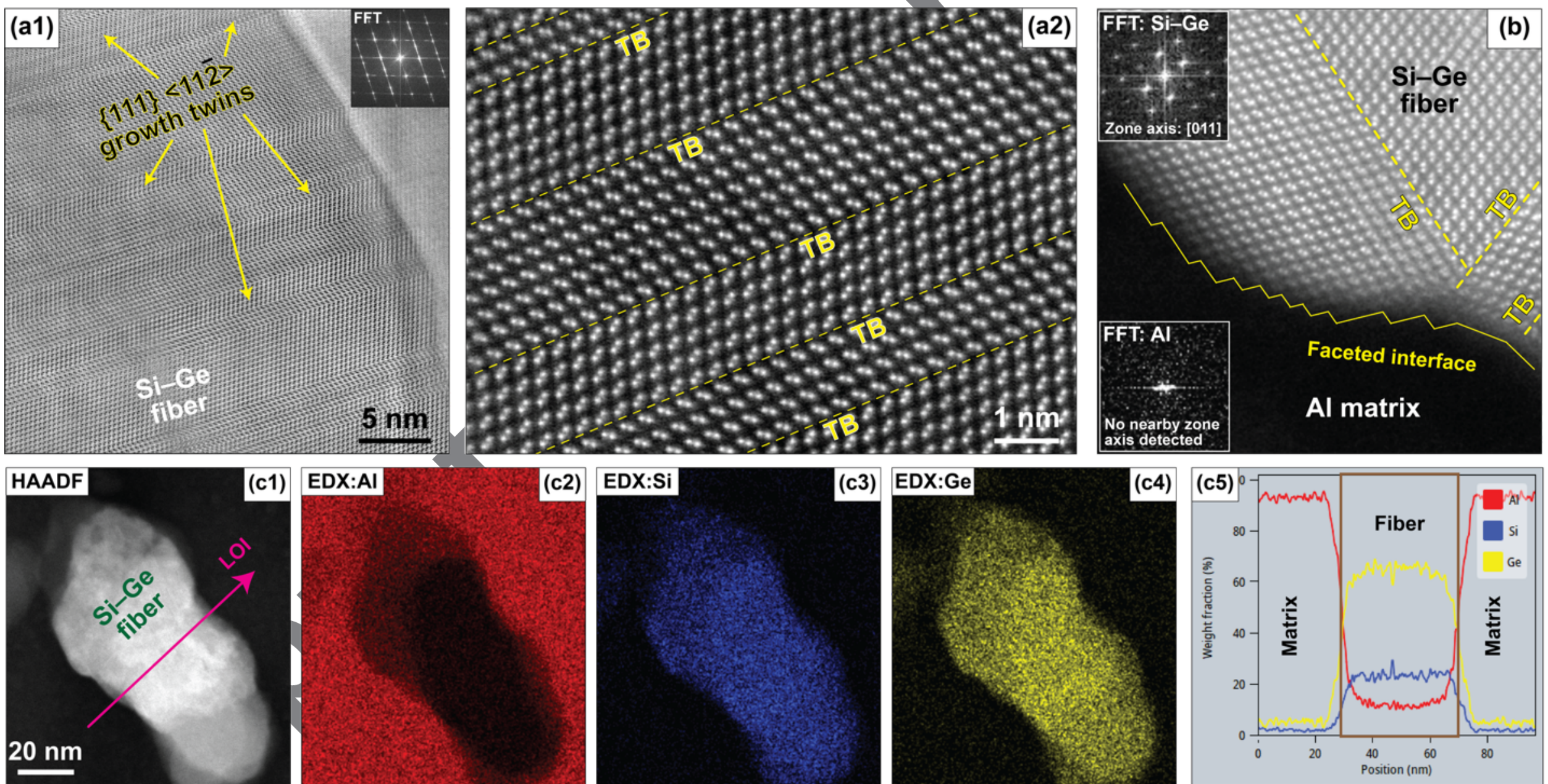

**Figure 5: Growth defects and Al retention in the Si- or (Si,Ge)-rich fibers.** (a1–a2) Representative HR-STEM images from eutectic D showing growth twins in (Si,Ge)-rich fibers; similar twin-containing fibers are observed across all compositions. (b) Atomic-resolution STEM-HAADF image showing a representative faceted Al/(Si,Ge) interface with terraces and ledges, the fiber is indexed as diamond-cubic (Si,Ge), while no reproducible Al/(Si,Ge) orientation relationship is established. (c1–c4) STEM-HAADF image and STEM-EDX elemental maps from composition D showing an (Si,Ge)-rich fiber embedded in the Al-rich matrix. (c5) Line profile taken along the indicated direction in (c1), showing Si and Ge enrichment within the fiber together with a finite retained Al signal.

### *(ii) Mechanical behavior assessed by uniaxial compression and tension tests*

The mechanical response of the laser-remelted eutectic composites was first evaluated by *in situ* SEM compression tests (refer to the Supplementary Videos 1–4). Representative cylindrical samples from compositions A–D had comparable dimensions, with diameters of ~8 μm and heights of ~18 μm (Figure 6a1–a4), enabling direct comparison of the compressive response across the alloy series. The true stress–true strain curves show that all compositions sustain stable plastic flow to strains up to ~30% without catastrophic load drops (Figure 6b). The yield strength decreases from $532 \pm 24$ MPa in the binary Al–Si reference eutectic A to $495 \pm 32$ MPa in composition B. This initial decrease is consistent with the increase in inter-fiber spacing and with the expectation that the flow stress for interface-confined slip decreases as the available slip distance in the Al-rich phase increases [7]. However, further Ge addition reverses this trend: compositions C and D exhibit higher yield strengths of $552 \pm 17$ MPa and $560 \pm 27$ MPa, respectively, despite their coarser eutectic spacings, indicating that Ge alloying introduces additional barriers to plastic flow beyond eutectic spacing (discussed in the next section).

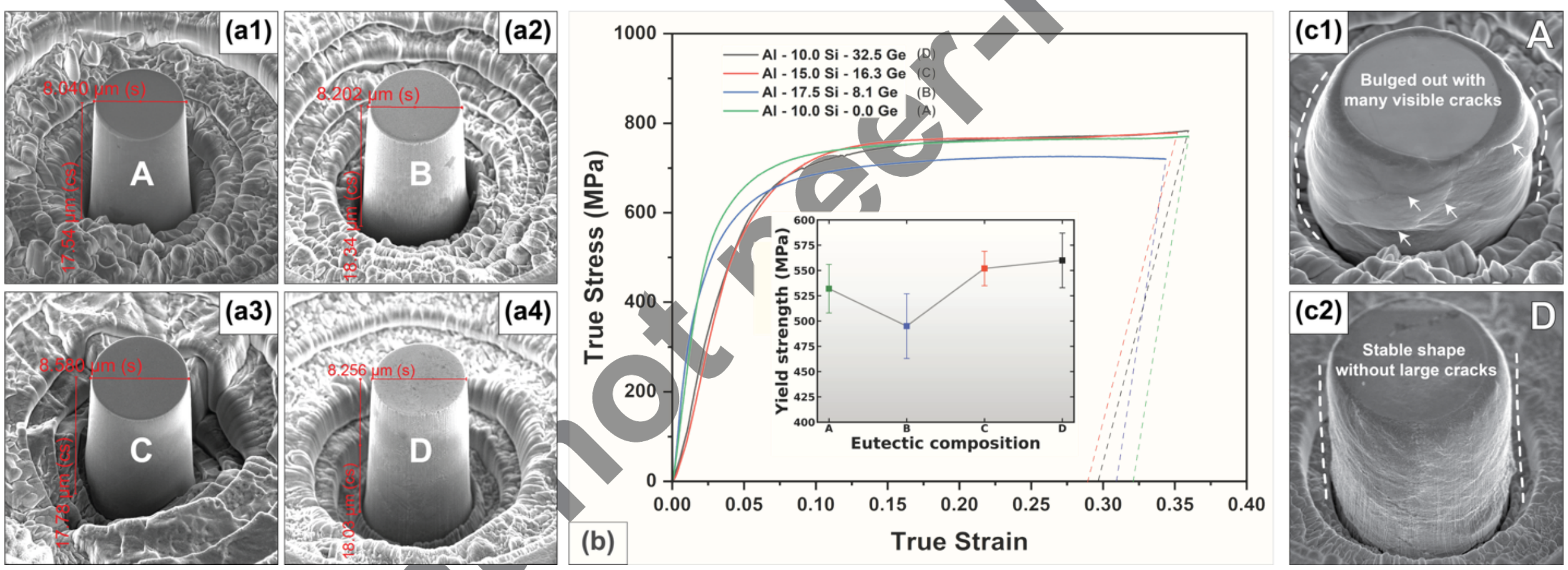

**Figure 6: Uniaxial compression response of laser-remelted eutectic composites.** (a1–a4) Representative as-fabricated micropillars from compositions A–D with comparable dimensions of ~8 μm diameter and ~18 μm height. (b) True stress–true strain curves from *in situ* SEM compression tests, with the inset showing the corresponding 0.2% offset yield strengths. All compositions sustain large plastic strains without catastrophic load drops, while the yield strength decreases from A to B and then increases for the higher-Ge compositions C and D. (c1–c2) Post-compression SEM images comparing deformation morphologies of eutectics A and D. The binary Al–Si eutectic A shows pronounced bulging and visible surface cracking, whereas the Ge-rich eutectic D retains a more stable pillar shape without large cracks after compression.

The post-compression sample morphologies further show that Ge addition improves deformation compatibility. The binary Al–Si pillar exhibits pronounced barreling/bulging and multiple visible surface cracks after compression (Figure 6c1), indicating strain localization and fracture. In contrast, the Ge-rich composition D pillar retains a more stable cylindrical shape without large surface cracks after comparable deformation (Figure 6c2). Thus, Ge addition does not simply retain strength despite coarsening; it also suppresses cracking during large plastic strain.

Although compression is well suited for identifying deformation mechanisms because the deformed volume remains compact and accessible for site-specific post-deformation TEM preparation, crack opening is mechanically suppressed, and hard-phase fracture may be partly masked by contact constraint, barreling, and frictional stress states under compression [54]. Tensile loading therefore provides a more stringent assessment of co-deformation because plastic incompatibility between the Al-rich matrix and the Si- or (Si,Ge)-rich fibers can directly promote interfacial decohesion, fiber cracking, and unstable damage propagation. Accordingly, *in situ* SEM micro-tension was performed to evaluate whether the Ge-containing eutectic also exhibits improved tensile stability and delayed damage accumulation under a crack-opening loading condition.

Figure 7a compares the true tensile stress–strain responses of the binary Al–Si reference eutectic A and the Ge-containing eutectic D. The initial linear slope of the stress–strain curves should not be interpreted as the elastic modulus because strain was inferred from gripper displacement rather than measured directly from the gauge section and therefore includes contributions from system compliance, local alignment, and gripping effects. Accordingly, the strain to failure is reported as the plastic strain after subtracting the apparent elastic/compliance contribution from the measured strain. Based on the engineering stress–strain response (refer to the Supplementary Figure S6), eutectics A and D exhibit comparable 0.2% offset yield strengths of approximately 550 MPa, while the engineering ultimate tensile strength increases from approximately ~680 MPa in A to ~710 MPa in D. The corresponding true stress–strain curves further show that eutectic D sustains a higher post-yield flow stress and reaches a maximum pre-necking true stress of approximately 790 MPa, compared with approximately 733 MPa for eutectic A. The true plastic strain to failure similarly increases from approximately 13% in A to approximately 15% in D. Thus, Ge addition improves post-yield strain accommodation and delays tensile instability despite producing a coarser eutectic microstructure. Notably, the Al–Si specimens exhibit discrete load drops during the early stages of plastic deformation. In situ SEM observations associate these events with the onset of localized Si-fiber cracking at approximately 2–3% plastic strain, indicating that damage accumulation begins substantially before final tensile failure.

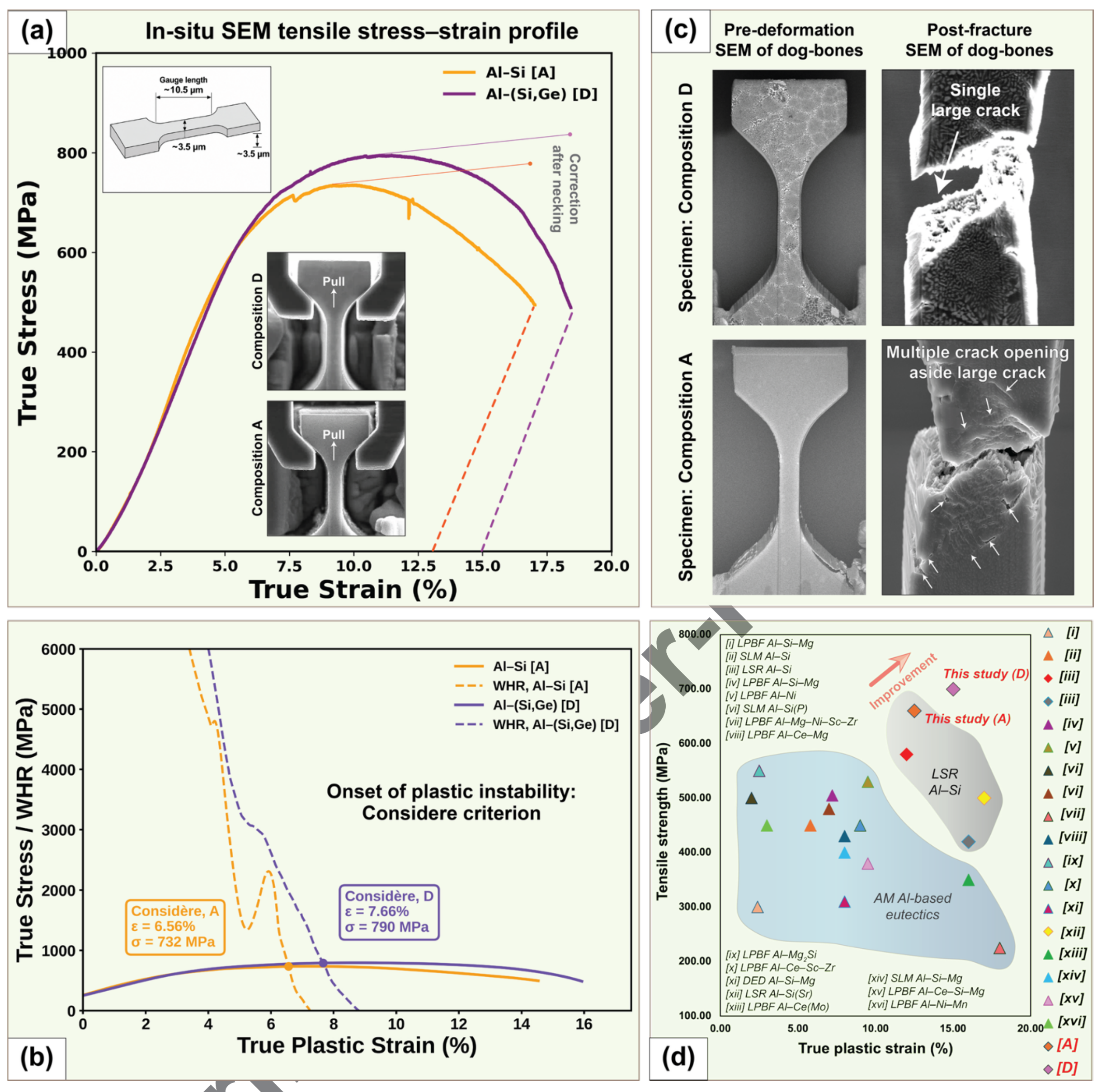

**Figure 7: Tensile behavior of the laser-remelted eutectic micro-dog-bone specimens.** (a) True stress–true strain curves of the binary Al–Si reference eutectic A and the Ge-containing eutectic D obtained by *in situ* SEM microtension. Insets show the micro-dog-bone geometry and representative SEM images during loading. (b) Work-hardening-rate (WHR) analysis showing the onset of tensile instability based on the Considère criterion; the crossover occurs at a larger strain for eutectic D than for eutectic A. (c) Pre-deformation and post-fracture SEM images of the tensile specimens. Eutectic D fails predominantly by a single dominant crack, whereas eutectic A exhibits multiple secondary crack openings in addition to the main fracture path. (d) Comparison of true tensile strength and true plastic strain of the present eutectics with literature reported Al-based eutectic and additively manufactured alloys, showing the improved strength–ductility balance of the Ge-containing eutectic.

The difference in tensile stability is further shown by the strain hardening analysis in Figure 7b. According to the Considère criterion, tensile instability begins when the work hardening rate (WHR) equals the true stress, i.e., $\frac{d\sigma}{d\varepsilon} = \sigma$ [55]. A comparative strain hardening analysis indicates that the apparent instability crossover occurs at a larger measured plastic strain in D than in A. This result is consistent with the higher flow stress and larger plastic strain sustained by eutectic D in Figure 7a. Post-fracture SEM images further support the improved tensile damage tolerance of the Ge-containing eutectic. Eutectic A exhibits multiple secondary crack openings in addition to the main fracture path, which is also reflected by load drops in the stress–strain curve. This distributed cracking indicates less effective accommodation of tensile incompatibility, with local stresses relieved through repeated crack nucleation. In contrast, eutectic D fails primarily through a dominant crack with fewer secondary crack openings before final failure, indicating more stable strain partitioning between the Al-rich matrix and the (Si,Ge)-rich fibers. Importantly, it indicates that the Ge-containing eutectic is more effective in delaying the transition from local incompatibility accommodation to unstable crack multiplication. It should be noted that the measured tensile plastic strain may still represent a lower-bound estimate of the intrinsic co-deformation capability of the eutectic microstructure. Because SEM-scale dog-bone specimens inevitably sample a limited number of eutectic colonies, colony boundaries cannot be completely avoided and may act as preferential sites for strain localization or crack initiation [56].

The property map in Figure 7d places the present Ge-containing eutectic in a favorable strength–ductility regime relative to reported Al-based laser-processed eutectic alloys, including laser powder bed fusion (LPBF), selective laser melting (SLM), direct energy deposition (DED), and laser surface remelting (LSR) systems [13, 56–70]. This comparison includes both *in situ* micro-tension and bulk tensile data from the literature and should therefore be interpreted with appropriate consideration of specimen size, testing geometry, and defect population. The tensile strength values in Figure 7d correspond to the ultimate tensile strength obtained from engineering stress–strain curves, whereas the true plastic strain values were determined from the true stress–strain response after subtracting the elastic strain contribution. For the present LSR melt pools, bulk-scale tensile testing is not feasible because the rapidly solidified eutectic region is spatially confined within the remelted zone; therefore, *in situ* SEM micro-tension provides the most direct assessment of the tensile response of the eutectic microstructure itself. Within this context, eutectic D occupies a region of concurrent high tensile strength and large plastic strain, outperforming the binary Al–Si reference and comparing favorably with reported laser-processed Al-based eutectic systems. This improvement is particularly notable because it is achieved despite the coarser eutectic spacing of composition D, further confirming that the tensile response too is not governed by length-scale refinement alone. This suggests that the hierarchical Ge-containing microstructure promotes more compatible strain partitioning between the Al-rich matrix and the (Si,Ge)-rich fibers, motivating detailed post-deformation microscopy to identify the underlying phase-specific deformation mechanisms.

### *(iii) Deformed microstructure*

The compression results show that the higher-Ge eutectics retain or recover yield strength despite coarser eutectic spacing and exhibit plastic flow to large strains. To identify the microstructural origin of this response, post-compression STEM analysis was first performed. In the binary Al–Si reference and low-Ge eutectic B, deformation in the Al-rich phase is primarily accommodated by arrays of single dislocations confined within narrow Al channels bounded by Si- or (Si,Ge)-rich fibers, as shown in the Supplementary Figure S7. This deformation mode agrees with prior studies of rapidly solidified Al–Si eutectics, where plasticity in the Al phase was shown to occur through confined channel slip and dislocation accumulation between closely spaced Si fibers [12, 13, 20]. In contrast, the higher-Ge compositions C and D exhibit deformation-induced boundary formation within the Al-rich channels. The region analyzed in Figure 8 was selected from the lower part of the compressed pillar (composition D), where finite element simulation (Appendix A) indicates a relatively lower local plastic strain compared with the highly deformed upper contact region. Although this region is still deformed well beyond yielding, the lower local strain increases the likelihood of preserving earlier-stage defect configurations before extensive dislocation accumulation and strain localization obscure the underlying mechanisms. Figure 8(a1and a3) displays a representative deformed Al channel subdivided into two adjacent regions, labeled sub-grain 1 and sub-grain 2, separated by a deformation-induced sub-boundary. After this subdivision, the effective slip distance in the Al-rich phase is reduced to ~50 nm, comparable to the original Al channel thickness in the binary Al–Si reference eutectic.

The two adjacent Al regions exhibit a small crystallographic misorientation. When each region is brought close to zone-axis orientation, BF-STEM imaging reveals arrays of individual dislocations within both sub-grain 1 and sub-grain 2 [Figure 8(a2 and a4)]. Thus, interface confined slip remains operative after Ge addition, but the relevant confinement length-scale is no longer defined by the original Al/(Si,Ge) inter-fiber spacing. Instead, in the higher-Ge eutectics, plastic deformation generates internal low-angle boundaries that subdivide the coarser Al channels, resulting in a smaller effective slip length. Atomic-resolution STEM-HAADF imaging further shows that the deformation-induced sub-boundary is a low-angle tilt boundary. Figure 8(b1–b2) shows two adjacent Al sub-grains separated by a boundary with a misorientation of ~6°. The boundary is composed of dislocation cores arranged into a wall-like configuration, characteristic of low-angle $\{011\}$-type tilt boundary formation by dislocation organization during plastic deformation. For an *fcc* Al dislocation with Burgers vector ($\vec{b} = \frac{a}{2}[011]$), a ~6° tilt misorientation corresponds to a dislocation spacing on the order of ~ 2 nm, matching the dense dislocation-wall contrast observed at the boundary. Thus, the Al-rich phase evolves from confined single-dislocation glide toward strain-induced subdivision as the local dislocation density increases. STEM-EDX analysis in Figure 8b1 shows enhanced Ge signal along the deformation-induced boundary, indicating Ge segregation to the dislocation boundary. The observation is significant because these boundaries form most prominently in the cluster-containing compositions C and D, where the Al-rich matrix already contains coherent (Si,Ge) clusters before deformation. Interestingly, the detectable density of such clusters in deformed Al matrix after boundary formation is much lower compared to the observed density prior to deformation in STEM-HAADF images.

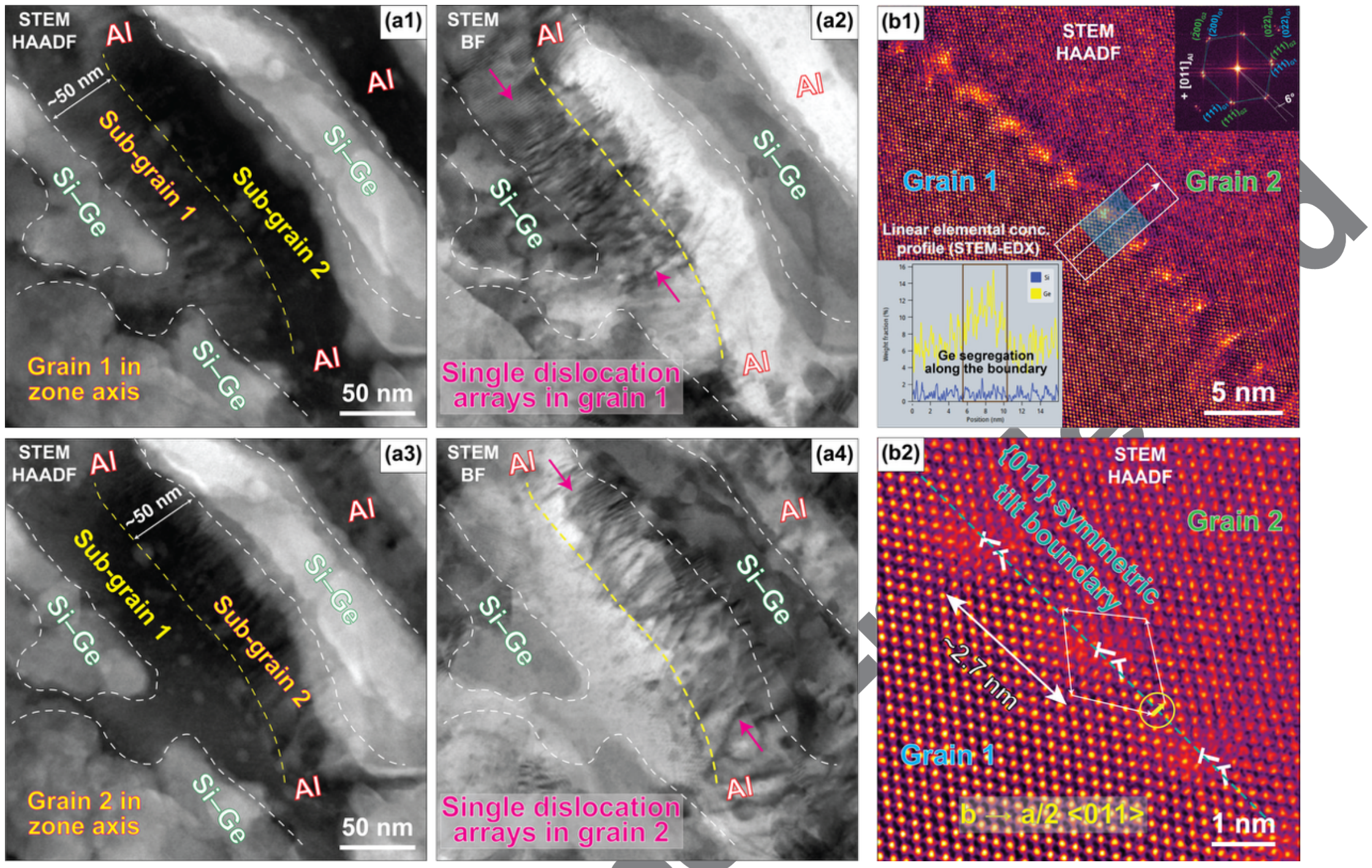

**Figure 8: Deformation-induced boundary formation and Ge segregation in the Al-rich phase after compression [from composition D].** (a1–a4) Post-compression STEM-HAADF and BF images showing deformation-induced subdivision of an Al-rich channel into two slightly misoriented regions (sub-grains), with confined dislocation arrays present in both regions. (b1–b2) Atomic-resolution STEM-HAADF images of a low-angle tilt boundary in the deformed Al phase, showing dislocation-wall formation and Ge enrichment along the boundary by STEM-EDX line profiling. The boundary formation introduces a deformation-generated confinement length scale within the coarsened Al channels.

After examining the early-stage deformation structure from the lower-strain region of the compressed pillar, we next analyze the Al-rich phase from the higher-strain region near the pillar top/contact surface and middle section, where deformation-induced subdivision of the Al-rich phase progresses beyond low-angle boundary formation and develops into a high-angle boundary network. Figure 9a shows a heavily deformed Al-rich region where dense dislocation networks develop within initially weakly misoriented Al domains due to activation of multiple slip systems. The accumulation and rearrangement of dislocations produce pronounced lattice rotation, indicating that confined channel slip evolves into progressive grain subdivision as deformation proceeds. When imaged from another specimen orientation with the (Si, Ge) fibers approximately aligned along the electron beam direction to reduce phase-overlap contrast, multiple deformation-induced boundaries extending across the Al phase are resolved [Figure 9b]. The Al phase is subdivided into multiple nanoscale grains between (Si, Ge) fibers, showing that deformation-induced boundary formation evolves from isolated channel subdivision to an extended boundary network at higher strain. Importantly, confined channel slip remains active even after this subdivision. Figure 9c shows BF-STEM contrast from subdivided Al regions, where dislocation arrays are still visible within individual grains bounded by (Si, Ge) fibers and neighboring deformation-induced boundaries. Thus, the deformation mode in the Al-rich phase is hierarchical: initial confined dislocation glide occurs within Al channels, followed by dislocation accumulation, low-angle boundary formation, progressive lattice rotation, and eventual high-angle (evident from Figure 9d1–d2) boundary development. This sequence generates new internal boundaries inside the originally coarser Al channels, further reducing the effective slip length for confined channel slip during deformation.

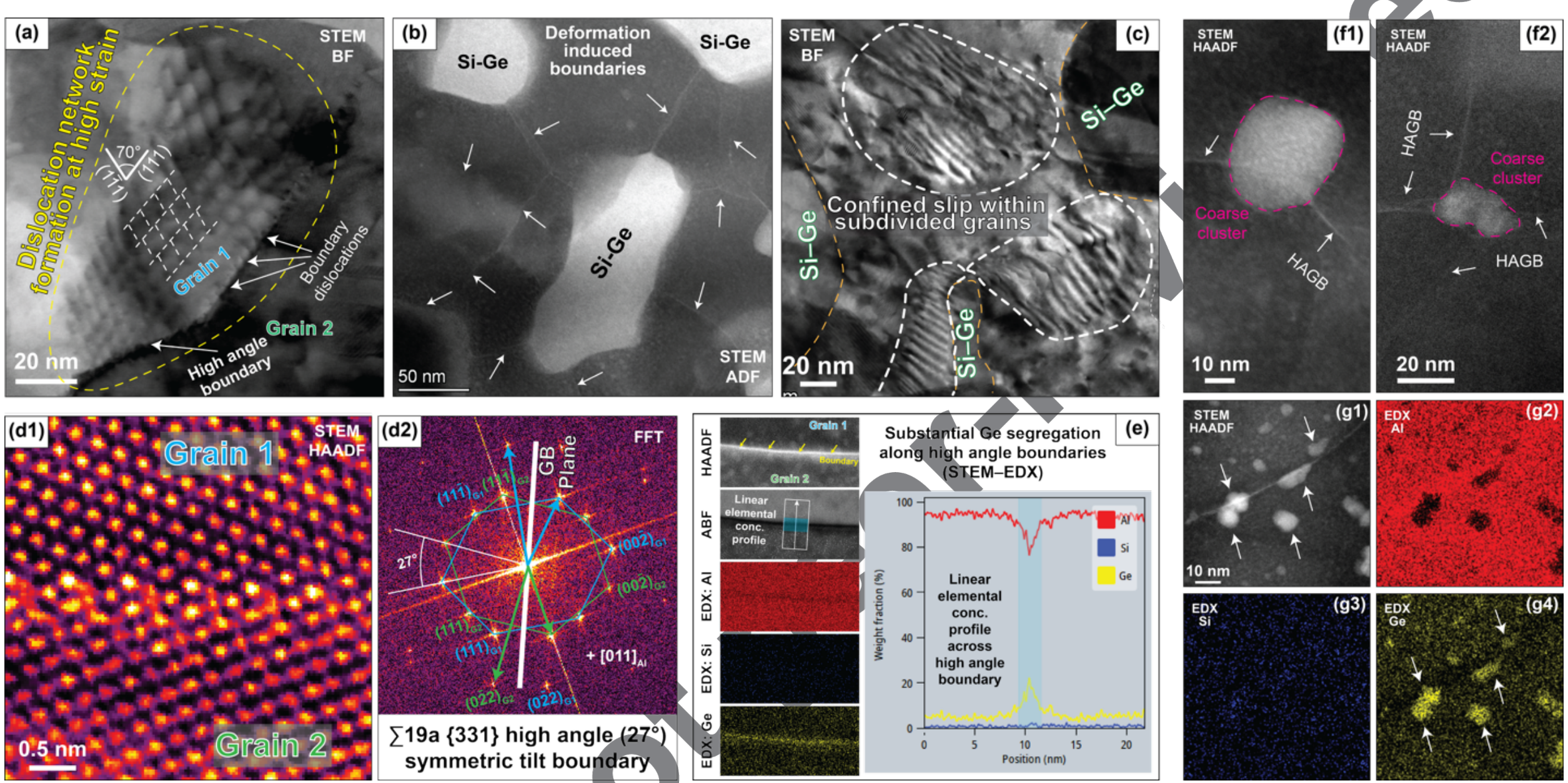

**Figure 9: High-angle boundary formation and Ge segregation in the Al-rich phase after compression.** (a) BF-STEM image from a higher-strain region of the compressed pillar showing dense dislocation networks and deformation-induced boundary formation in the Al-rich phase. (b) ADF-STEM image acquired with fibers approximately aligned with the electron beam, revealing an extended high-angle boundary network within the confined Al-rich phase. (c) BF-STEM image showing that confined channel slip remains active within subdivided Al regions after boundary formation. (d1–d2) Atomic-resolution STEM-HAADF image and corresponding FFT analysis of a representative deformation-induced high-angle boundary, indexed as a $\sum 19a\ \{331\}$-type symmetric tilt boundary with a misorientation of ~27°. (e) STEM-EDX maps and line profile showing preferential Ge segregation along the high-angle boundary. (f1–f2) STEM-HAADF images showing coarse post-deformation clusters spatially associated with a HAGB and a HAGB triple junction, respectively. (g1–g4) STEM-HAADF image and corresponding Al, Si, and Ge EDX maps of representative coarse clusters, demonstrating that these clusters are predominantly Ge-rich with little detectable Si. The occurrence of Ge-rich coarse clusters along many, though not all, deformation-induced boundaries and triple junctions, together with Ge segregation along the boundaries themselves, is consistent with localized deformation-assisted redistribution of Ge within the Al-rich phase.

High resolution imaging and local diffraction analysis in Figure 9(d1–d2) revealed that deformation-induced boundaries have evolved into high-angle boundaries with increasing plastic strain. The STEM-HAADF image shows two adjacent Al grains separated by a sharp boundary, and the corresponding FFT contains two distinct sets of *fcc* Al reflections. Both diffraction patterns are viewed along the common $[011]_{Al}$ zone axis, while the reflections from Grain 2 are rotated by approximately 27° relative to those from Grain 1. Because the two grains share a common $\langle 011\rangle$ axis, this in-plane rotation corresponds directly to a tilt misorientation about $[011]_{Al}$. The boundary plane is found as $\{331\}$, the resultant of $\{111\}$ and $\{220\}$ planar peaks, and can be indexed as $\sum 19a$ $\{331\}\langle 110\rangle$ tilt boundary. For a cubic $\langle 110\rangle$ high-angle symmetric tilt boundary with a boundary plane belonging to the $\{hhl\}$ family, the ideal misorientation angle is given by [71]:

$$\theta = 2\tan^{-1}\left(\frac{l}{h\sqrt{2}}\right) \quad (1)$$

For the $\{331\}$ boundary, *h*=3 and *l*=1, giving: $\theta = 26.53°$. The experimentally measured rotation of approximately 27° is therefore in good agreement with the ideal 26.53° misorientation expected for a $\sum 19a$ $\{331\}\langle 110\rangle$ symmetric tilt boundary. The simultaneous agreement among the common $\langle 011\rangle$ tilt axis, the measured angular separation of the two diffraction patterns, and the indexed $\{331\}$-type boundary plane supports the crystallographic assignment. The identification of this boundary as a $\sum 19a$ $\{331\}\langle 110\rangle$ symmetric tilt boundary is also physically reasonable because this boundary type has been commonly reported as a high-angle CSL boundary in *fcc* Cu [72, 73].

It should be noted that the deformation-induced high-angle boundary network is not expected to consist exclusively of this boundary character. Because the Al-rich phase undergoes three-dimensional lattice rotation during compression, many adjacent grains cannot be simultaneously brought close to the same zone axis in a thin TEM foil. In such cases, one grain may be well aligned for atomic-resolution imaging while the neighboring grain is significantly off-zone, making reliable boundary-plane and misorientation indexing difficult. An example of such a non-indexable high-angle boundary is provided in the Supplementary Figure S8. Therefore, the $\{331\}$-type assignment should be interpreted as the reproducible crystallographic character of the subset of high-angle boundaries that could be imaged with both adjoining grains near a common $[011]$ zone axis, rather than as an exhaustive description of all deformation-induced high-angle boundaries. Within this experimentally accessible subset, $\{331\}$-type boundaries were repeatedly observed.

STEM-HAADF micrograph along with EDX analysis further reveals pronounced Ge enrichment along the high-angle boundaries [Figure 9(e)]. The elemental maps and line profile show a localized increase in Ge concentration at the boundary, reaching values as high as ~20 wt.% Ge, accompanied by a corresponding decrease in Al. This enrichment is substantially greater than that measured at the low-angle boundaries (~10 wt.% Ge). In contrast, no comparable Si enrichment is detected. The preferential accumulation of Ge therefore demonstrates that chemical redistribution accompanies the deformation-induced subdivision of the Al-rich phase. Additional evidence of this redistribution is provided by the coarse (average diameter ranging 5–20 nm) solute-rich features observed in the vicinity of the high-angle boundary network. Figure 9f1 shows a representative coarse cluster located directly along a high-angle boundary, while Figure 9f2 identifies another coarse cluster associated with a junction between several high-angle boundaries. Such coarse clusters were frequently observed along or near deformation-induced boundaries and boundary junctions, although they were not present at every boundary. Their spatial association with the boundary network is therefore nonuniform rather than ubiquitous. The chemistry of these coarse clusters is revealed in Figure 9(g1–g4) using STEM-EDX as predominantly Ge-rich, rather than retaining the correlated (Si,Ge) enrichment characteristic of the finer clusters observed in the as-processed Al-rich phase. These post-deformation Ge-rich clusters are also substantially coarser than the (Si,Ge) clusters present in the as-processed condition, indicating a marked change in both cluster size and chemistry following deformation. The similar chemical selectivity of the coarse clusters and deformation-induced boundaries – both showing pronounced Ge enrichment with little detectable Si – indicates that deformation is accompanied by a spatial redistribution of Ge within the Al-rich phase. The frequent occurrence of Ge-rich coarse clusters along the boundary network and particularly at some boundary junctions further suggests a close spatial relationship between this redistribution and the deformation-generated defect structure.

We next investigate whether the (Si,Ge) fibers also participate in plastic deformation. This is important because, in conventional Al–Si eutectics, the hard *dc* phase often acts as a brittle reinforcement and accommodates strain primarily through cracking or fragmentation [20, 74], which is also verified for the reference eutectic A in this study (refer to Supplementary Figure S9). In the present Ge-containing eutectics, however, post-compression microscopy shows that the (Si,Ge) fibers remain largely intact and develop deformation-induced planar defects instead of catastrophic fracture. Figure 10 compares the (Si,Ge) fibers before and after compression in eutectic D. The low-magnification STEM-HAADF images in Figure 10(a1–a2) indicates that the overall fibrous eutectic morphology is retained after deformation, with no obvious signs of fiber cracking. This supports the improved post-compression sample morphology shown earlier for composition D and indicates that the (Si,Ge) phase does not simply fail by brittle fracture during large plastic strain. Higher-magnification ADF-STEM images from same [011] zone axis further reveal the change in internal defect structure of the fibers. Before deformation, the (Si,Ge) fibers contain long growth twins that extend across much of the fiber width, reflecting TPRE-mediated faceted eutectic growth [Figure 10b1]. After compression, additional planar faults appear within the fibers [Figure 10b2]. These faults are characteristically shorter than the growth twins, are often confined to local regions near Al/(Si,Ge) interfaces or pre-existing twin boundaries, and do not extend continuously across the entire fiber. Their short length scale, localized distribution, and absence in the corresponding pre-deformation fibers indicate that they are deformation-induced rather than growth defects. The tendency of these faults to initiate or terminate near interfaces and twin boundaries further suggests that they form through partial-dislocation activity at sites of local stress concentration owing to strong elastic/plastic incompatibility between the confined shearing Al channels and the stiff *dc* fibers.

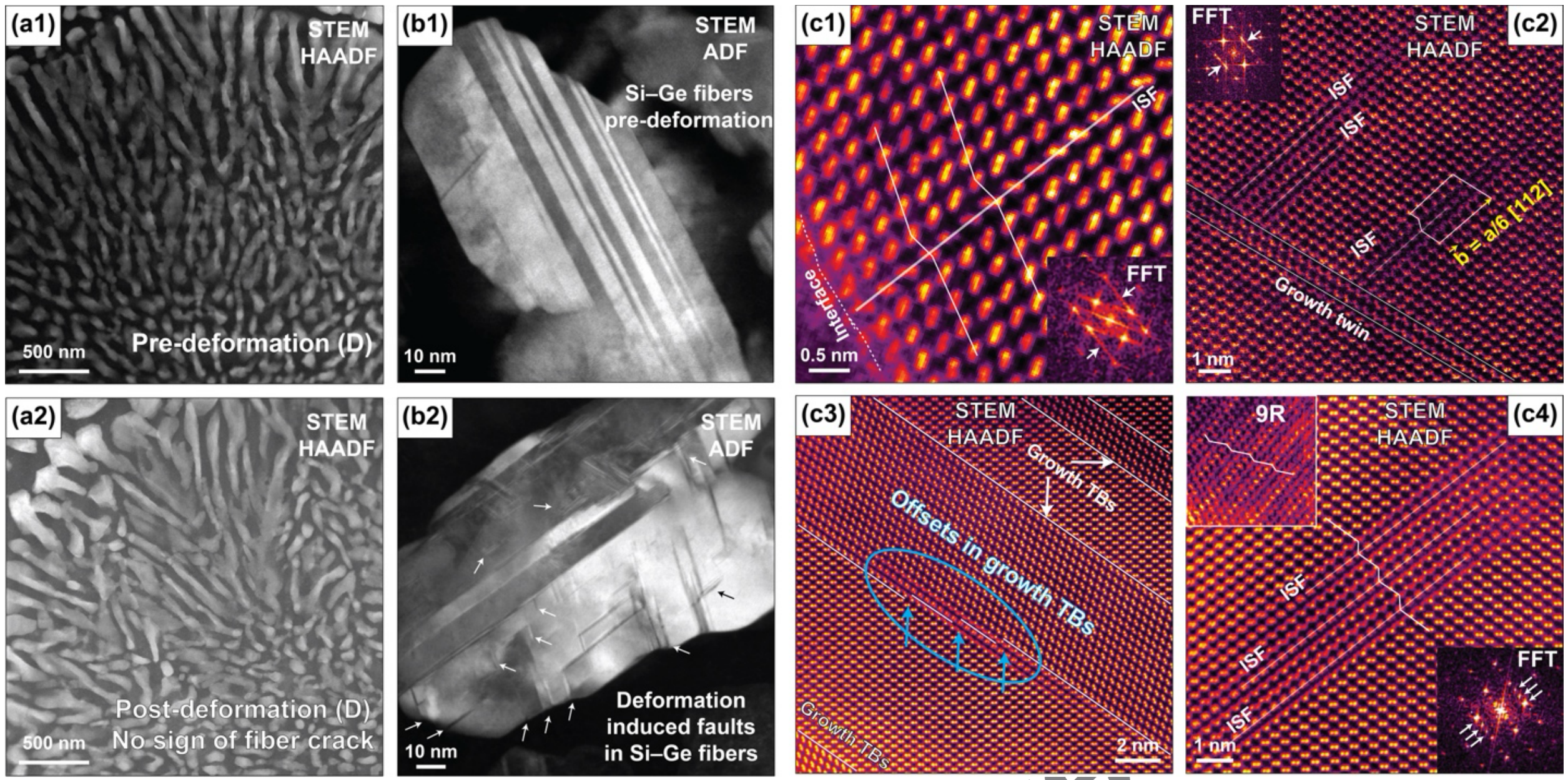

**Figure 10: Deformation-induced planar faulting in (Si,Ge) fibers after compression.** (a1–a2) STEM-HAADF images of eutectic D before and after compression, showing that the fibrous (Si,Ge) morphology is retained after deformation without obvious widespread fiber cracking. (b1–b2) ADF-STEM images comparing the internal defect structure of (Si,Ge) fibers before and after deformation. Long growth twins are present before deformation, whereas shorter, localized planar faults appear after compression, indicating deformation-induced fault formation. (c1–c4) Atomic-resolution STEM-HAADF images showing intrinsic stacking faults on {111} planes and their activity near pre-existing growth twins and Al/(Si,Ge) interfaces. FFT patterns in the inset highlight streaking in planar peaks due to the pointed planar faults. The observed fault geometry is compatible with Shockley-partial-dislocation activity in the diamond-cubic (Si,Ge) phase.

Atomic-resolution STEM-HAADF images in Figures 10(c1–c4) confirm the crystallographic nature of these planar defects. The faults lie on close-packed $\{111\}$ planes of the *dc* (Si,Ge) phase. In the as-processed state, the dominant planar features are long growth twins and occasional growth-induced stacking faults. After deformation, the density of shorter faulted regions increases, and multiple intrinsic stacking-fault segments are observed within individual fibers at potential stress concentration sites like interfaces [Figure 10c1] and growth twin boundaries [Figure 10c2]. Sometimes, offsets/steps on growth twin boundaries have been recorded as a result of local shearing [Figure 10c3]. Although occasional, even formation of 9R-like localized structures were captured where there is presence of stacking faults in every third $\{111\}$ atomic plane [Figure 10c4]. These features are compatible with Shockley partial-dislocation activity with Burgers vectors of type $\vec{b} = \frac{a}{6}\langle 112\rangle$. Such partials may either nucleate directly under the high local stresses generated near Al/(Si,Ge) interfaces and twin boundaries, or form through dissociation of perfect dislocations according to [75]:

$$\frac{a}{2}[1\bar{1}0] \longrightarrow \frac{a}{6}[2\bar{1}\bar{1}] + SF + \frac{a}{6}[1\bar{2}1] \quad (2)$$

where *SF* denotes the stacking fault left between the leading and trailing Shockley partials. Therefore, the observed intrinsic fault segments indicate that the (Si,Ge) fibers accommodate local strain through partial-dislocation-mediated planar faulting rather than cracking. The observed *post-mortem* fault density may therefore represent a lower bound on the total partial-dislocation activity during loading, because some leading partials may have partially retracted during unloading; direct *in situ* TEM deformation would be required to quantify the dynamic nucleation, glide, and recovery of partial dislocations in the (Si,Ge) fibers [76, 77].

## 4. Discussion

The central finding of this work is that laser rapid solidified Al–(Si,Ge) eutectics exhibit higher strength and ductility than binary Al–Si, in spite of Ge alloying coarsening the eutectic microstructures. In this section, the mechanistic origins of this behavior are discussed by considering how Ge-induced clustering, solute segregation, deformation-induced boundary formation, and partial-dislocation-mediated faulting collectively enable plastic co-deformation of the Al-rich matrix and (Si,Ge)-rich fibers.

### (i) Strengthening beyond eutectic spacing: transition from geometric confinement to chemistry-assisted confinement

Interface confined slip model provides a useful approximation of yield strength because plastic yielding is expected to initiate in the softer Al-rich phase. In this framework, the flow stress of the confined Al matrix can be estimated by considering the combined contributions from geometrically necessary dislocations (GNDs) and interface confined slip (ICS):

$$\sigma_m = \left[\alpha' G b \sqrt{\left\{\frac{2\sqrt{3}\varepsilon_p}{bh}\right\}}\right]_{GND} + \left[M \frac{Gb}{8\pi h'} \frac{(4-\upsilon)}{(1-\upsilon)} \left\{ln \frac{\alpha h'}{b}\right\}\right]_{ICS} \quad (3)$$

where (M) is the Taylor factor of Al, taken as ~3. The remaining parameters were chosen following prior work [20, 78, 79]:

$G$ (shear modulus of Al) = 26.1 GPa, $b$ (Burgers vector) = 0.286 nm (calculated for full dislocations on $\{1\ 1\ 1\}_{Al}$), $\upsilon$ (Poisson's ratio) = 0.34, $\alpha$ (core cut-off parameter) = 0.6. The single-dislocation length ($h$) can be estimated from Figure 2f (measured edge-to-edge fiber spacing is assumed equivalent to Al channel thickness) and the projected dislocation length on the slip plane, ($h'$), is approximated as ($h$). In the present alloys, GNDs primarily arise from the interfaces between soft Al matrix and rigid (Si,Ge) fibers. $\alpha'$ is taken as 0.2 and 0.4 for cluster free and cluster containing Al phases [61]. For the yield strength estimate, $\varepsilon_p$ was taken as 0.002, corresponding to the 0.2% offset condition. If $h$ is taken as the as-solidified inter-fiber spacing or Al channel width, this model predicts that the yield strength should decrease monotonically as the eutectic spacing increases from A to D. This trend is observed only from A to B. The strength recovery in C and D therefore indicates that the relevant slip length during deformation is not simply the original inter-fiber spacing. Instead, the post-compression STEM results show that the higher-Ge eutectics develop internal low- and high-angle boundaries within the Al-rich phase. Because the reported yield strength is defined using the 0.2% offset criterion, a finite amount of plastic strain and associated dislocation accumulation has already occurred at this point. Under the strong nanoscale confinement imposed by the surrounding (Si,Ge)-rich fibers, such early dislocation storage may initiate dislocation-wall formation and incipient low-angle subdivision, thereby reducing the operative slip distance below the original as-solidified Al-channel width. These deformation-induced boundaries subdivide the coarser Al channels and reduce the operative slip distance to a smaller effective length scale. At lower strain, low-angle boundary formation is observed in Figure 8, which subdivides the coarser Al channels. After subdivision, the effective thickness available for confined slip is reduced to ~50 nm in composition D, comparable to the Al channel thickness in composition A. Although the exact subdivision length at the 0.2% offset condition cannot be directly determined from the postmortem specimens, the ~50 nm sub-grain dimension observed in the lower-strain region of composition D is used here as a representative estimate of the reduced operative slip length once subdivision begins. Using this approach, the calculated strength contribution from Al matrix ($\sigma_m$) for composition A, where clusters are absent, and composition D, where clusters are prominent, are ~480 MPa and ~540 MPa, respectively.

However, the experimentally measured yield strength is a composite-level response rather than the stress carried by the Al-rich phase alone. At the onset of yielding, the Al-rich matrix is expected to yield first, whereas the Si- or (Si,Ge) fibers primarily contribute through elastic load sharing. Therefore, the composite yield strength can be approximated as:

$$\sigma_y = \sigma_m V_m + \sigma_f V_f \tag{4}$$

where $\sigma_m$ and $\sigma_f$ present the stress contribution from the matrix and fibers, and $V_m$ (=1– $V_f$) and $V_f$ are their respective volume fractions that have been estimated and reported in Figure 2(a–d). Since the fibers deform predominantly elastically during initial deformation due to strong plastic incompatibility, $\sigma_{Cr}$ can be approximated as:

$$\sigma_f = E_f \times \varepsilon \tag{5}$$

For pure Si phase, elastic modulus ($E_{Si}$) is often approximated as 170 GPa [80]. However, for (Si,Ge) phases as in composition D, elastic modulus ($E_f$) of several thousands of randomly oriented fibers (with respect to loading direction) can be estimated following the average of Voigt (upper bound) and Reuss (lower bound) rules of mixture, respectively given as [81, 82]:

$$E_f^{Voigt} = E_{Si} f_{Si} + E_{Ge} f_{Ge} \tag{6}$$

$$\frac{1}{E_f^{Reuss}} = \frac{f_{Si}}{E_{Si}} + \frac{f_{Ge}}{E_{Ge}} \tag{7}$$

At 0.2% offset strain, approximated as the elastic strain carried by the fiber at the onset of composite yield ($\varepsilon$), for composition D (fraction of Si in (Si,Ge) fibers, $f_{Si}$~0.45; fraction of Ge in (Si,Ge) fibers, $f_{Ge}$~0.55), and considering $E_{Ge}$ ~ *130 GPa* [83], $E_f$ becomes ~147 GPa. Thus, $\sigma_f$ for composition A and D will be ~350 MPa and ~300 MPa, respectively. Replacing $\sigma_m$, $\sigma_f$, $V_m$ (=1– $V_f$) and $V_f$ with their determined values for both eutectics A and D in *Equation (4)* yield nearly similar yield strengths of about 465 MPa.

Eutectic colony boundaries, i.e., Al–Al grain boundaries, are likely secondary strengthening sources since the micropillars sample multiple such colonies. Because the confined-slip calculation does not explicitly include the baseline resistance associated with colony-scale Al–Al grain boundaries, a Hall–Petch-type offset was estimated using the eutectic colony size:

$$\sigma_{H-P} = \sigma_0 + kd^{-1/2} \quad (8)$$

For pure Al (*fcc*), the friction stress ($\sigma_0$) and Hall-Petch coefficient ($k$) are often approximated as ~20 MPa and 0.10 MPa.m$^{1/2}$, respectively [84, 85], which results in $\sigma_{H-P} \approx 75$ MPa for an average eutectic colony size of ~5 μm [Figure 2(a, d)]. Note that the eutectic colony size is significantly larger than the eutectic inter-fiber spacing, so the estimated strengthening contribution from Al–Al grain boundaries is relatively smaller. Adding up this colony boundary strengthening term with the prior calculated yield strength for a single colony results in effective yield strengths of ~540 MPa both for eutectic D and reference eutectic A, which is in reasonable agreement with the experimental values.

The analysis above, particularly the confined-slip-strengthening estimate for the Al-rich matrix, shows that the retained strength of the higher-Ge eutectics requires an internal deformation-generated confinement length scale within the Al-rich phase. The key question is why such deformation-induced boundaries form and persist more readily in the higher-Ge compositions. This behavior can be understood by considering the effects of pre-existing (Si,Ge)-rich clusters and deformation-induced Ge segregation in stabilizing the evolving dislocation-boundary structure. Solid-solution hardening contributions from Ge and Si to dislocation glide in Al were not computed separately since the solutes (particularly Ge) primarily influenced the evolution of dislocation sub-boundaries in Al that refined the effective slip distance in interface-confined slip.

### *(ii)* ***Role of (Si,Ge)-rich clusters and Ge segregation in stabilizing deformation-induced sub-boundaries in Al***

The deformation-induced boundaries observed within the Al-rich phase indicate that plasticity in the present eutectic is not governed only by the as-solidified eutectic spacing. Instead, the Al-rich channels undergo a coupled structural and chemical reorganization during deformation. In the as-solidified condition, the Al-rich matrix contains nanoscale (Si,Ge)-rich clusters, while the surrounding (Si,Ge)-rich fibers impose strong geometric confinement. During compression, plastic strain is primarily accommodated by confined dislocation slip in the Al-rich phase. Because the adjacent (Si,Ge)-rich fibers cannot accommodate the same plastic strain through equivalent deformation modes, dislocations accumulate near Al/(Si,Ge) interfaces, fiber terminations, cluster-rich regions, and local variations in channel geometry. These local dislocation pile-ups and strain gradients promote lattice rotation and the formation of deformation-induced low-angle boundaries. With continued plastic strain, further dislocation absorption and divergent rotation of neighboring Al regions can transform such dislocation walls into high-angle boundaries, consistent with deformation-induced grain subdivision reported in plastically deformed *fcc* metals [86,87].

The important distinction in the present Al–(Si,Ge) eutectic is that this grain subdivision occurs inside a chemically heterogeneous Al-rich matrix. Therefore, the deformation-induced boundaries are not only structural products of dislocation accumulation; they also act as chemical sinks within a supersaturated, cluster-containing matrix. Prior studies on severely deformed Al alloys have shown that plastic deformation can simultaneously modify defect structure and redistribute solute: precipitates or clusters may be sheared, partially dissolved, and redistributed as dislocations interact with them [88, 89]. In such cases, gliding dislocations can serve as short-circuit diffusion paths that absorb and transport solute atoms, while evolving dislocation boundaries and grain boundaries become preferential sites for solute accumulation [89]. Deformation-induced non-equilibrium segregation can also be described by vacancy–solute-complex models, in which plastic deformation generates excess vacancies, vacancy–solute complexes form locally, and gradients in the mobile defect–solute population drive solute transport toward boundary sinks [90].

This framework provides a physically plausible route for Ge redistribution in the present eutectic without requiring long-range lattice diffusion through defect-free Al. The nanoscale (Si,Ge)-rich clusters in the Al-rich matrix provide a local solute reservoir, while deformation-induced dislocation walls and high-angle boundaries provide nearby sinks. The local transport (flux) can be expressed as [90, 91]:

$$J_{Ge} = -D_{eff} \nabla C_{Ge}^{*} \quad (9)$$

where $C_{Ge}^{*}$ represents the mobile Ge-containing solute/defect population and $D_{eff}$ is an effective transport coefficient that includes vacancy-assisted diffusion, dislocation-pipe diffusion, and boundary-assisted transport. Because the characteristic distance between clusters and evolving boundaries is on the order of the nanoscale Al channel dimensions, only short-range redistribution is required. Thus, the Ge enrichment observed after deformation is best interpreted as local defect-assisted redistribution rather than conventional bulk diffusion.

The post-deformation chemistry further indicates that this redistribution is selective. Before deformation, the Al-rich matrix contains randomly distributed (Si,Ge)-rich clusters. After deformation, however, the deformation-induced Al boundaries and boundary junctions are enriched primarily in Ge, with little detectable Si enrichment. Therefore, the original (Si,Ge) cluster chemistry is not simply transferred intact to the boundaries. Instead, the coupled (Si,Ge) solute heterogeneity appears to chemically decouple during plastic deformation. Si may remain diffusely distributed in the Al-rich matrix, persist in residual cluster remnants below the spatial or chemical sensitivity of STEM-EDX. Ge, in contrast, preferentially partitions to the newly formed

Al boundaries, suggesting a stronger thermodynamic and/or kinetic affinity for the defect-rich boundary structure.

This selective Ge enrichment is mechanistically important because deformation-induced boundaries in Al are not automatically stable. They form because confined plasticity stores dislocations and generates lattice rotation, but such boundaries may also migrate, recover, or annihilate if the energetic penalty of maintaining them remains high. Solute segregation can reduce this penalty by relaxing the local boundary structure, lowering the effective boundary energy, and exerting a drag force on residual boundary dislocations, or disconnections. Room-temperature deformation-induced solute segregation has been directly observed at dislocation cores and faceted twin boundaries in Mg–Y, where solute enrichment is energetically associated with local lattice distortion and kinetically assisted by defect–solute interactions during plastic deformation [92]. Similarly, solute segregation along deformation-induced kink, twin-like, and tilt boundaries in Mg–Zn–Y has been proposed to lower interfacial energy and pin boundary motion [93]. Recent work on additively manufactured alloys also shows that solute-segregated dislocation structures can dynamically interact with gliding dislocations to sustain dislocation storage and plastic flow [94].

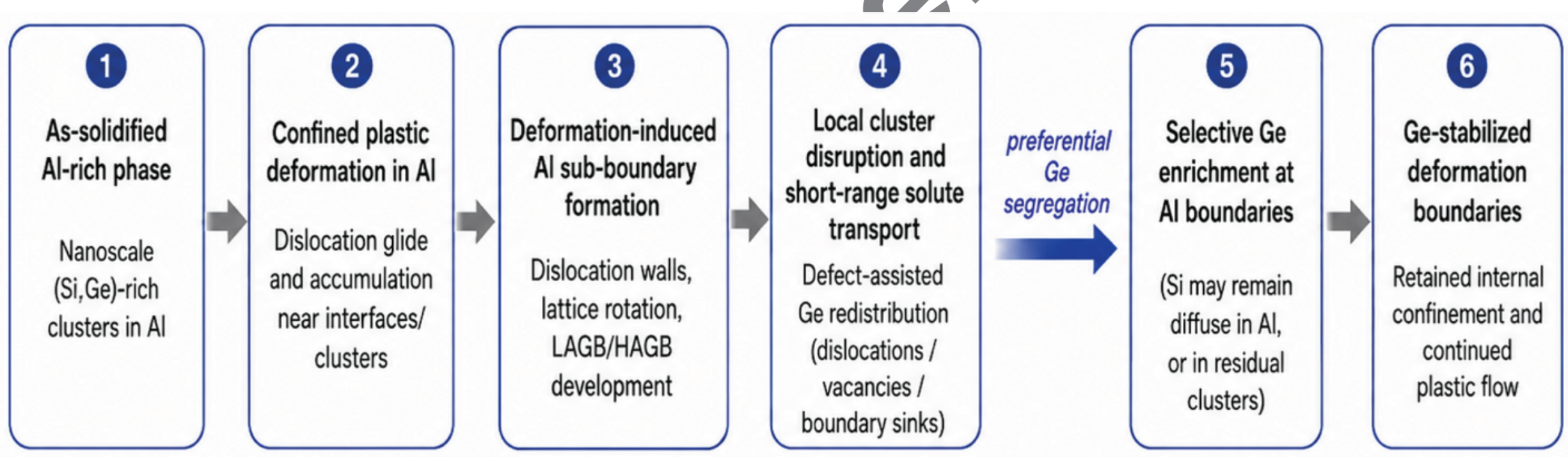

**Figure 11: Flow diagram of Ge-assisted deformation mechanism.** Proposed defect-assisted Ge redistribution and boundary-stabilization sequence in the Al-rich phase.

By analogy, Ge segregation in the present eutectic is proposed to stabilize the deformation-induced Al boundaries after they form. This does not imply that Ge alone creates the high-angle boundaries or prescribes the observed boundary character. The primary driving force for boundary formation remains the stored energy associated with confined dislocation plasticity in the nanoscale Al-rich channels. The role of Ge is instead to reduce the energetic and kinetic cost of retaining the newly formed boundaries. In this way, Ge converts otherwise transient dislocation walls into more stable internal interfaces, preserving a reduced effective slip length even though the primary eutectic spacing becomes coarser with increasing Ge content. The proposed evolution of Ge-assisted deformation mechanism in Al has been summarized using a flow diagram in Figure 11.

To test whether Ge can thermodynamically stabilize such deformation-induced boundaries, DFT calculations were performed on the experimentally observed $\sum 19a$ $\{331\}\langle 110\rangle$ high-angle Al boundary. Figure 12a shows the modeled boundary and the investigated Ge substitution sites near the boundary plane. The calculated segregation energies reveal a strong site dependence, with four of the five boundary-region sites being energetically favored relative to a bulk-like Al site. The screening segregation energies [95],

$$E_{seg}^{i} = E_{Ge@site\ i} - E_{Ge@bulk}, \qquad (10)$$

where $E_{Ge@site\ i}$ and $E_{Ge@bulk}$ are the total energies of otherwise identical $Al_{73}Ge_1$ supercells containing Ge at the grain-boundary and bulk-like sites, respectively] were approximately −0.341, −0.059, −0.107, −0.283, and +0.029 eV for GB1–GB5 [Figure 12b], respectively. The strongest segregation preference occurs at GB1, located directly within the boundary core. Importantly, the segregation energy does not vary monotonically with distance from the nominal boundary plane; for example, GB4 shows a stronger segregation tendency than the geometrically closer GB2 and GB3 sites. This indicates that Ge segregation is controlled primarily by the local atomic configuration of the boundary, rather than simply by distance from the boundary plane.

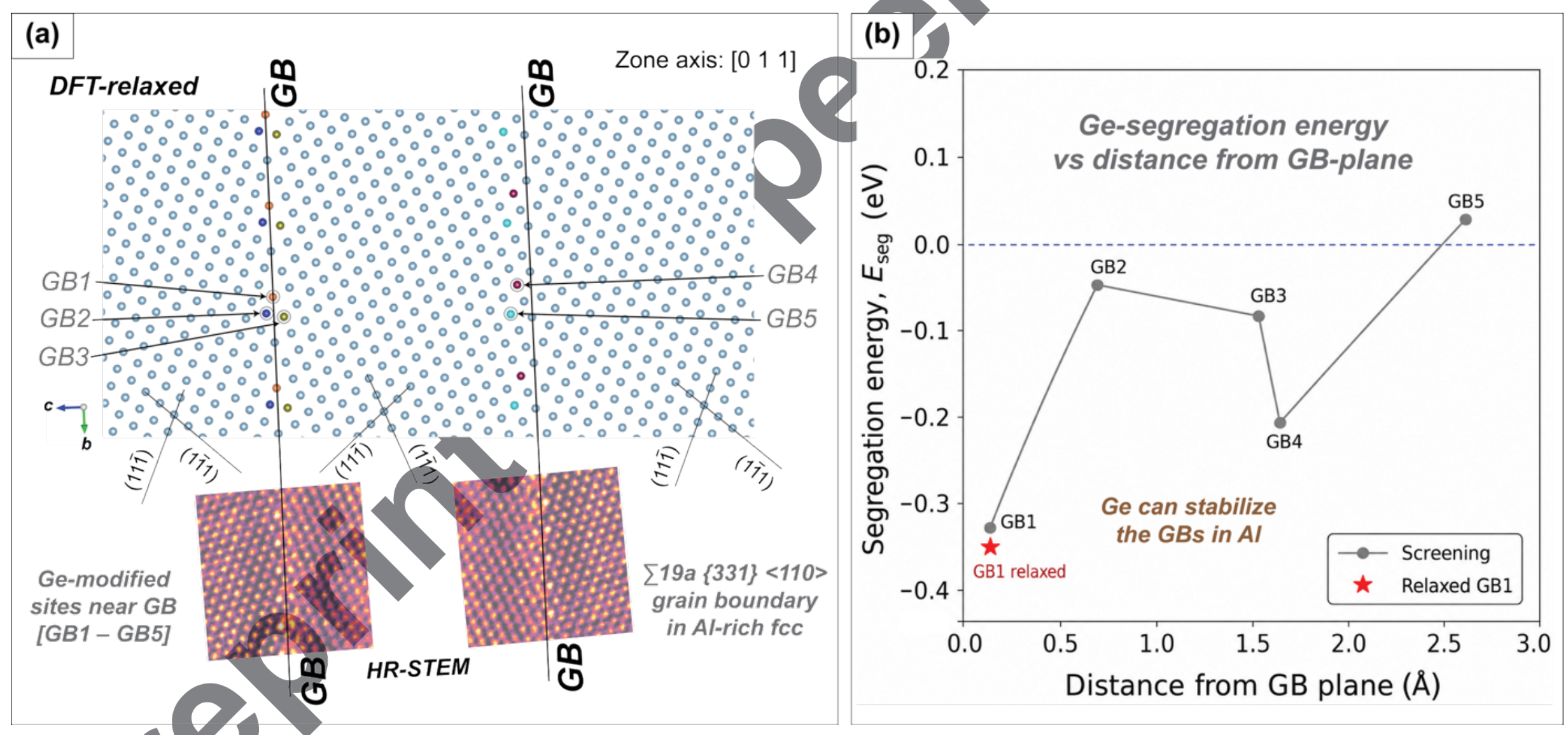

Figure 12: Density-functional-theory assessment of Ge segregation to the deformation-induced Al grain boundary. (a) Atomistic representation of the $\sum 19a$ $\{331\}\langle 110\rangle$ symmetric tilt boundary in *fcc* Al, showing the investigated Ge substitution sites near the grain-boundary plane, denoted GB1–GB5. Representative relaxed grain-boundary structures are shown below. (b) Calculated Ge segregation energy as a function of distance from the nominal grain-boundary plane. Negative segregation energies indicate preferential Ge occupation of the boundary region relative to a bulk-like Al site. Four of the five investigated sites are energetically favored, with the strongest preference occurring at the GB-core site GB1. Full structural relaxation of GB1 retains a strong negative segregation energy of approximately −0.349 eV, supporting a thermodynamic preference for Ge occupation of favorable sites within the deformation-induced Al boundary.

Full structural relaxation of the most favorable GB1 configuration retains a strong negative segregation energy of approximately −0.349 eV, close to the fixed-geometry screening value. This negative relaxed segregation energy demonstrates a substantial thermodynamic preference for Ge occupation of favorable atomic sites within the boundary core relative to substitution within the grain interior. The DFT results therefore provide atomistic support for the experimentally observed Ge enrichment at deformation-induced high-angle boundaries in the Al-rich phase. Mechanistically, these calculations support the interpretation that confined plastic deformation first generates dislocation walls and high-angle boundary structures, after which Ge preferentially redistributes toward favorable boundary-core sites. The thermodynamic preference for Ge occupation of favorable boundary-core sites can therefore lower the energetic cost associated with the segregated boundary structure and may promote its retention during continued deformation. Importantly, the calculations demonstrate preferential Ge segregation and energetic stabilization of an existing Al boundary; they do not imply that Ge alone nucleates the boundary. Although direct Si-decorated boundary calculations would be required to fully quantify the thermodynamic origin of the experimentally observed Ge-over-Si selectivity, prior first-principles studies show that Ge–vacancy binding in Al is stronger than Si–vacancy binding, supporting the possibility that deformation-generated vacancies and boundary defects preferentially assist Ge redistribution relative to Si [96, 97].

### ***(iii) <u>Plasticity in the (Si,Ge) fibers</u>***

The second contribution to improved deformation compatibility arises from localized plastic accommodation within the (Si,Ge)-rich fibers. In the binary Al–Si reference, the *dc* Si phase remains intrinsically resistant to room-temperature plastic flow and is consequently susceptible to cracking or fragmentation [13, 74, 98]. Nevertheless, deformation-induced stacking faults are already observed locally within the Si fibers of eutectic A, particularly near Al/Si interfaces, indicating that partial-mediated deformation can be activated when sufficiently large local stresses develop. With increasing Ge content, this accommodation becomes progressively more persistent: limited cracking remains evident in B, whereas the (Si,Ge) fibers in C and D remain largely intact at substantially higher strains and contain numerous short deformation-induced planar faults (Supplementary Figure S9). Thus, the observed response should not be interpreted as homogeneous plastic flow of the covalent fiber phase, but rather as localized defect-mediated strain accommodation at highly stressed sites.

The mechanical environment of these fibers differs fundamentally from that of bulk Si or bulk (Si,Ge). The hard *dc* phase is continuously embedded within a much softer Al-rich matrix that undergoes extensive plastic shear. Elastic–plastic incompatibility and load transfer across the Al/(Si,Ge) interfaces therefore generate large local stresses within the confined fibers, particularly at the interfaces and twin-boundaries. Direct crystallographic slip transmission from Al into (Si,Ge) is unlikely to dominate because no reproducible Al/(Si,Ge)orientation relationship is observed [99]. Dislocations undergoing confined slip in the Al-rich phase can therefore be accumulated at poorly aligned interfaces, producing intense local stress concentrations capable of activating defects within the adjacent (Si,Ge) phase. At the same time, the nanoscale fiber dimensions and high density of Al/(Si,Ge) interfaces strongly limit the characteristic distance available for crack extension, favoring crack arrest and redistribution of the local stress field. Consequently, cleavage and localized plastic accommodation need not be mutually exclusive within the hard phase. This high-stress condition is particularly important for room temperature plasticity in covalent semiconductors at fine length-scales. Chen *et al.* reported partial dislocation activity in Si nanopillars [100]. Dodson and Tsao showed experimentally that conventional bulk testing of Si-based semiconductors is generally limited by brittle fracture before the strongly stress-dependent regime of dislocation motion can be accessed; using strained (Si,Ge) structures, they extended the resolved shear stress to 2.6 GPa and demonstrated that the glide activation energy becomes strongly stress dependent at high stress [101]. Their analysis further showed that models neglecting the stress dependence of glide activation energy fail in this regime. This provides an important analogy for the present ultrafine eutectic: plastic flow of the surrounding Al phase can continuously load the nanoscale (Si,Ge) fibers while the dense interface network limits catastrophic crack propagation, potentially allowing the hard phase to access a defect-mediated deformation regime that is difficult to sustain in bulk Si or (Si,Ge).

The chemical state of the fibers further modifies this competition. Prior analysis confirms retention of Al within both the Si- and (Si,Ge)-rich fibers, and the calculations in Figure 13a1 indicate that dilute Al slightly reduces the intrinsic resistance of *dc* Si to crystallographic shear. For a $Si_{63}Al$ configuration containing ~1.56 at.% Al, homogeneous shear along the $(111)[1\bar{1}0]$ slip system produces a consistently lower resolved shear stress than in pure Si, with a reduction of approximately 0.18–0.22 GPa over the investigated strain range. The corresponding shear-induced energy,

$$\Delta E(\gamma) = E(\gamma) - E(0), \tag{11}$$

is also systematically lower for $Si_{63}Al$ [Figure 13a2]. Similar first-principles homogeneous-shear approaches have been employed to evaluate the intrinsic shear response of Si [102]. These calculations do not represent explicit dislocation motion and therefore should not be interpreted as a direct measurement of Peierls stress or dislocation mobility. Rather, they indicate that the local bond distortion required to accommodate slip-related shear becomes slightly less energetically and mechanically costly in the presence of retained Al. Under the high local stresses generated within the eutectic, this perturbation of the covalent bonding network may assist the initial activation of defect-mediated accommodation.

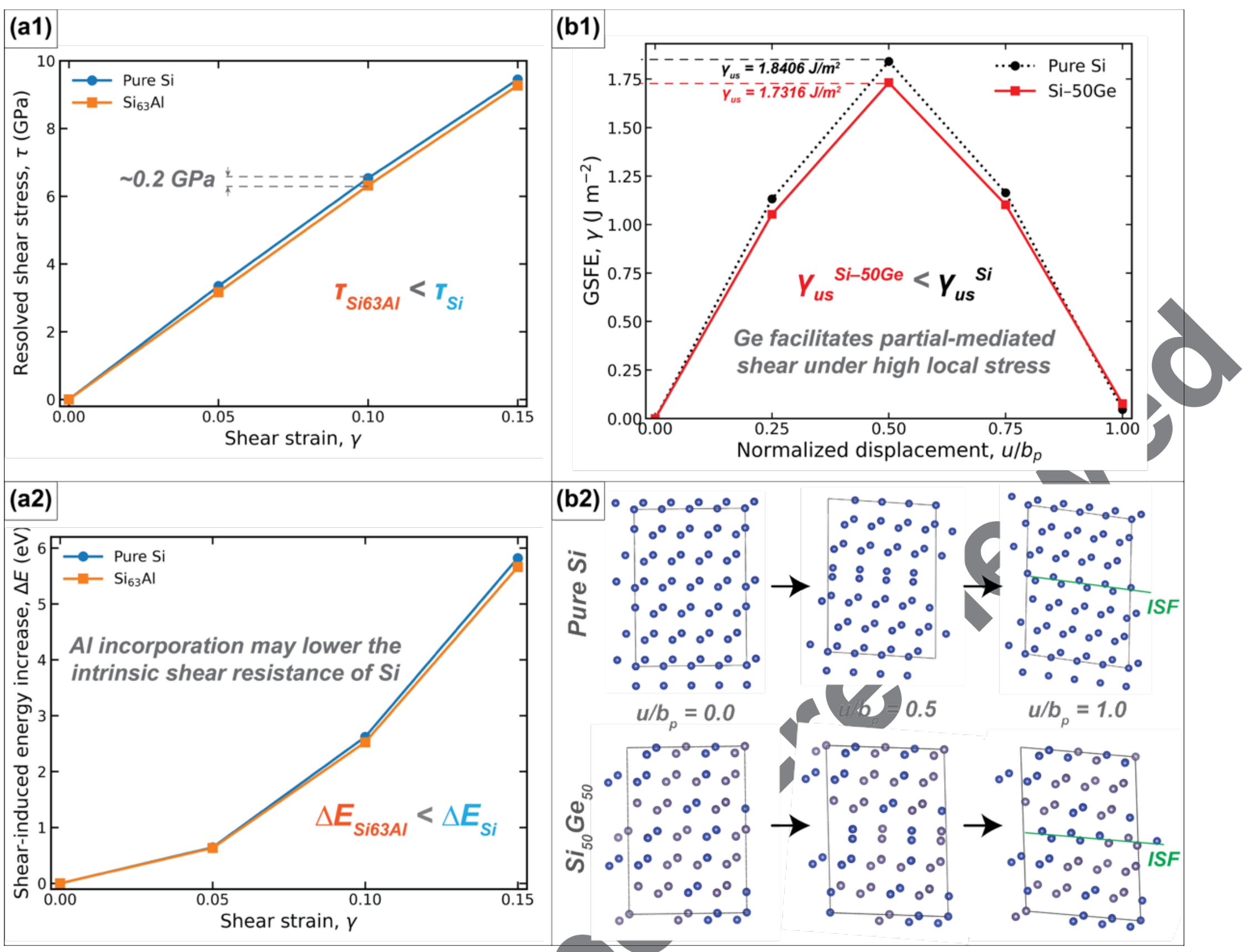

Figure 13: First-principles assessment of chemistry-assisted shear in the Si-based fibers. (a1) Resolved shear stress and (a2) shear-induced energy of pure Si and $Si_{63}Al$, showing reduced resistance to lattice shear with dilute Al incorporation. (b1) Generalized stacking fault energy profiles for pure Si and representative Si–50Ge along the $(111)$ glide-set $^{a}/_{6}\langle 112\rangle$ Shockley-partial pathway, showing a ~6% reduction in $\gamma_{us}$ with Ge. (b2) Corresponding relaxed configurations $\frac{u}{b_p} = 0$, 0.5, and 1.0.

Ge introduces a complementary effect on the partial-dislocation pathway itself. The generalized stacking-fault calculations in Figure 13(b1–b2) were performed along the $(111)$ glide-set $^{a}/_{6}\langle 112\rangle$ Shockley-partial pathway for pure Si and a representative Si–50Ge configuration. The generalized fault energy was evaluated as [103]

$$\gamma(u) = \frac{E(u) - E(0)}{A} \qquad (12)$$

where $u$ is the imposed shear displacement and $A$ is the $(111)$ fault-plane area. Both compositions exhibit the maximum energy near $\frac{u}{b_p} = 0.5$; however, the unstable GSFE decreases from $\gamma_{us}^{Si} = 1.8406\ J/m^2$ to $\gamma_{us}^{Si-50Ge} = 1.7316\ J/m^2$, corresponding to a reduction of approximately 6%. The Si–50Ge curve also remains systematically below that of pure Si at intermediate displacements surrounding the unstable configuration. Importantly, the energy of the final intrinsic stacking fault configuration does not decrease with Ge addition. Ge therefore does not simply stabilize stacking faults; rather, it lowers the unstable energetic barrier that must be traversed during leading

Shockley-partial glide. This interpretation is consistent with previous studies showing that the influence of Ge on dislocation activity in (Si,Ge) is strongly dependent on stress state and dislocation character, rather than reflecting a universal reduction in lattice resistance. Iunin *et al*. reported that Ge addition increased the stress required to initiate 60° dislocation motion, while enhancing dislocation velocity at sufficiently high applied stress [104]. Vanderschaeve and Caillard further demonstrated that, under high-stress conditions, Shockley-partial mobility is strongly controlled by dislocation-core structure, Peierls friction, and metastable dissociation configurations [105]. Calculations for shuffle screw dislocations similarly show that the Peierls barrier does not decrease monotonically from Si to Ge and is sensitive to the assumed core structure and $\gamma$-surface [106]. Accordingly, the present reduction in $\gamma_{us}$ should be interpreted specifically as a lowering of the energetic barrier along the glide-set Shockley-partial shear pathway considered here, rather than as a general softening of (Si,Ge).

The evolution across A–D can therefore be understood as a cooperative deformation mechanism. Plastic shear of the surrounding Al-rich matrix generates the high local stresses required to activate deformation within the hard fibers, while nanoscale confinement and the dense interface network restrict catastrophic crack propagation. Retained Al slightly lowers the energetic cost of lattice shear, whereas Ge further reduces the unstable barrier for leading Shockley-partial motion. Thus, Ge does not render bulk (Si,Ge) intrinsically ductile; rather, under the extreme stress state imposed by the ultrafine eutectic, it shifts the local competition from cleavage toward fault-mediated strain accommodation, as revealed by increasing persistence of stacking faults and suppression of fiber cracking from A to D.

### ***(iv) Chemistry-assisted co-deformation mechanism***

The mechanisms discussed above indicate chemistry-assisted co-deformation rather than independent deformation of the two constituent phases. In the Al-rich phase, confined slip produces dislocation storage and lattice subdivision, which evolve into low- and high-angle deformation-induced boundaries with increasing strain. Selective Ge redistribution to these boundaries provides a chemical contribution to retention of the subdivided structure. The negative segregation energies obtained from DFT show that Ge occupation of favorable boundary-core sites is thermodynamically preferred relative to a bulk-like Al site, providing an energetic basis for the experimentally observed boundary segregation. Boundary formation therefore preserves a reduced operative slip length within the coarsened Al channels without terminating confined slip; instead, plastic flow continues within the newly subdivided Al regions through single dislocation glide, and Ge-assisted boundary stabilization. Compared with binary Al–Si, where confined slip is more readily exhausted by dislocation accumulation and Si cracking, the progressive formation and Ge-assisted retention of internal boundaries provide additional sites for dislocation storage, rearrangement, and continued strain accommodation. In the (Si,Ge)-rich fibers, chemistry assists deformation through a complementary DFT-supported mechanism. Al retention lowers the calculated resistance to lattice shear in *dc* Si, while Ge incorporation reduces the unstable generalized stacking-fault-energy barrier along the glide-set Shockley-partial pathway. These effects do not imply homogeneous plasticity of the covalent fiber phase, but they lower the barrier for localized defect-mediated accommodation under the high local stresses generated by Al/(Si,Ge) incompatibility. Thus, the Al-rich phase sustains plastic flow through Ge-stabilized internal subdivision, while the (Si,Ge)-rich phase locally accommodates incompatibility through partial-dislocation-mediated planar faulting rather than widespread brittle cracking. This coupled response explains how Ge addition retains strength and improves damage tolerance despite eutectic coarsening, as schematically summarized in Figure 14.

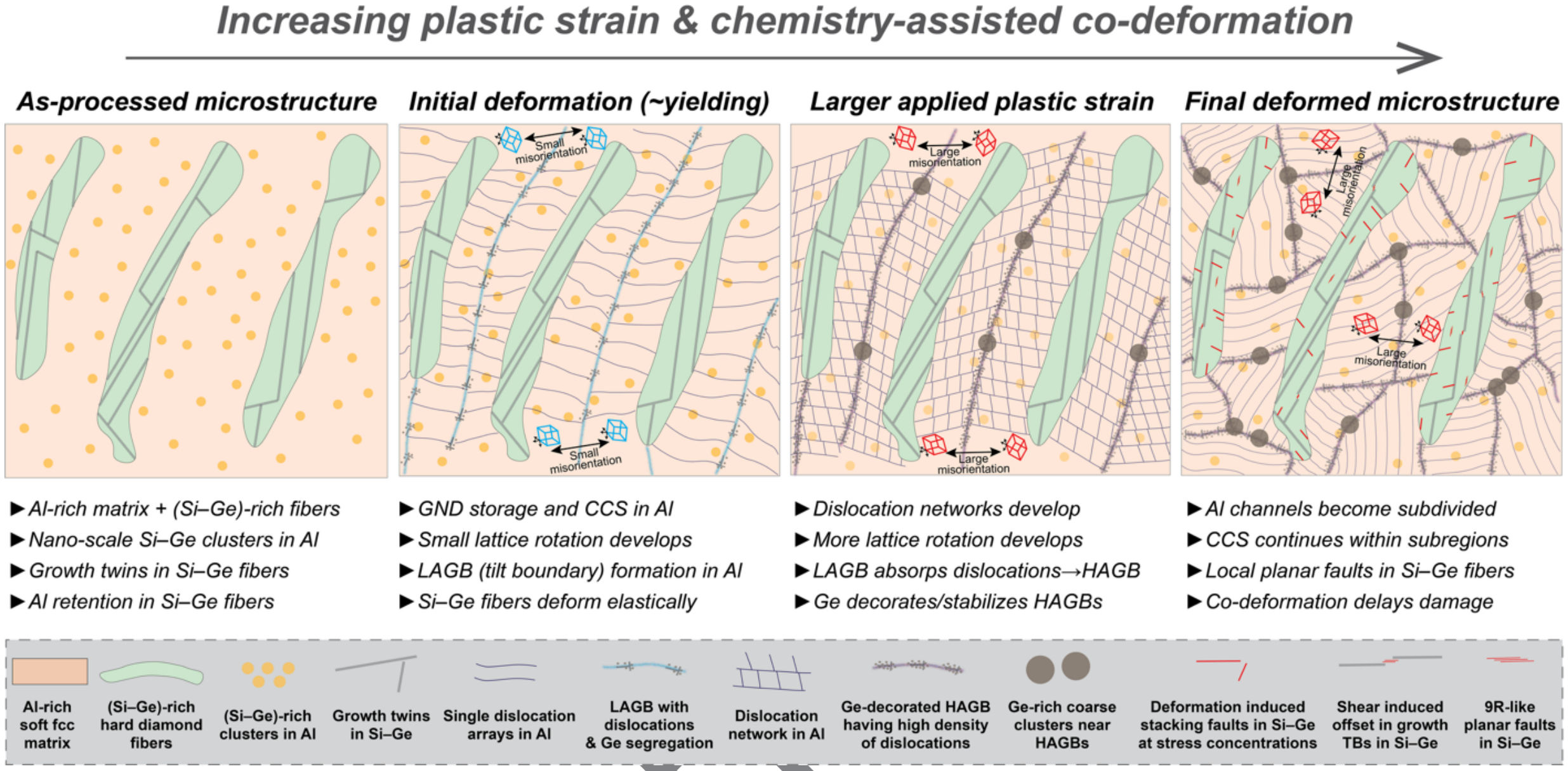

**Figure 14: Overall deformation behavior.** Schematic illustration of deformation mechanisms in the Al–(Si,Ge) eutectic with increasing plastic strain.

## 5. *Summary and conclusions*

This work demonstrates hierarchical eutectic design strategy for enabling plastic co-deformation between metallic and hard (covalent) phases. The key conclusions are given below:

- Laser rapid solidification preserved the desired two-phase Al/(Si,Ge) nano-scale fibrous morphology across all the univariant ternary eutectic compositions (A–D) without forming separate Ge-rich or intermetallic phases. Increasing Ge content coarsened the eutectic length scale, with average fiber diameter increasing from ~32 nm in A to ~51 nm in D and average inter-fiber spacing increasing from ~52 nm to ~98 nm, while the fiber-phase fraction increased to ~33% in D from ~15% in A. Several hierarchical features, including faceted interfaces, growth twins, and Al retention within the fiber phase, are inherent to the rapidly solidified eutectic morphology, whereas increasing Ge content promotes (Si,Ge)-rich clustering in the Al-rich phase and increases Ge incorporation within the diamond-cubic fibers.
- The compressive yield strength was retained or enhanced despite eutectic coarsening. The 0.2% offset yield strength initially decreased from $532 \pm 24$ MPa in eutectic A to $495 \pm 32$

MPa in eutectic B, consistent with increased inter-fiber spacing. However, further Ge addition increased the yield strength to $552 \pm 17$ MPa in C and $560 \pm 27$ MPa in D, demonstrating that strengthening is not controlled by the primary eutectic spacing alone. Tensile testing confirmed improved post-yield stability and delayed damage localization. Eutectics A and D exhibited comparable 0.2% offset tensile yield strengths of ~550 MPa. The engineering UTS increased from ~680 MPa in A to ~710 MPa in D. The corresponding maximum pre-necking true stress increased from ~733 to ~790 MPa, while the true plastic strain to failure increased from ~13% to ~15%. Early load drop is consistent with premature fiber fracture in eutectic A. Work hardening analysis further showed a delayed Considère instability in D, while post-fracture SEM revealed fewer secondary crack openings, confirming improved tensile compatibility.

- The Al-rich phase accommodated plasticity through a transition from confined slip to deformation-induced grain subdivision. In the binary and low-Ge eutectics, deformation was dominated by confined dislocation glide within Al channels. In the higher-Ge eutectics, dislocation storage and lattice rotation produced deformation-induced low- and high-angle boundaries inside the Al-rich phase. These boundaries subdivided the coarsened Al channels and restored a smaller effective slip length, allowing confined slip to remain active even after microstructural coarsening.
- Ge preferentially segregated to deformation-induced boundaries in Al and is proposed to assist their retention. STEM-EDX revealed Ge enrichment along deformation-induced low- and high-angle boundaries, including reproducibly indexed $\sum 19a\ \{331\}\langle 110\rangle$-type boundaries among crystallographically analyzable cases. DFT calculations for the experimentally observed $\sum 19a$ boundary showed negative Ge segregation energies at four of five investigated boundary-region sites, demonstrating a thermodynamic preference for Ge occupation of favorable boundary-core environments. These results support a mechanism in which confined dislocation storage provides the primary driving force for grain subdivision, while subsequent Ge segregation lowers the energetic cost of solute occupation at the newly generated boundaries and may assist retention of the deformation-generated substructure.
- The (Si,Ge)-rich fibers participated locally in strain accommodation through planar-fault-mediated deformation. Unlike the Si-rich fibers in the binary Al–Si reference, which eventually cracked at higher strain, the Ge-containing fibers remained largely intact and developed deformation-induced stacking faults, twin-boundary offsets, and occasional 9R-like faulted regions. These defects are consistent with localized Shockley partial-dislocation activity at stress concentration sites like faceted interfaces, and twin-boundaries, rather than homogeneous plasticity of the covalent phase. DFT calculations confirm Al retention lowers the calculated resistance to lattice shear in diamond-cubic Si, while Ge incorporation reduces the unstable generalized stacking-fault-energy barrier along the glide-set Shockley-partial pathway, favoring stacking fault mediated plasticity.

- Overall, Ge addition enables chemistry-assisted co-deformation rather than simple length-scale-dependent strengthening often reported in ultrafine eutectic heterostructures. This mechanism produces a strength–ductility synergy by retaining high strength while delaying hard-phase fracture and associated microstructural damage.

# Supplementary material: Chemical and interface confinement effects in promoting plastic co-deformation in high-strength nano-scale eutectics

Arkajit Ghosh [#], Amit Misra [#]

Department of Materials Science and Engineering, University of Michigan – Ann Arbor, MI 48109, USA

[#] Corresponding author: arkajitg@umich.edu, amitmis@umich.edu

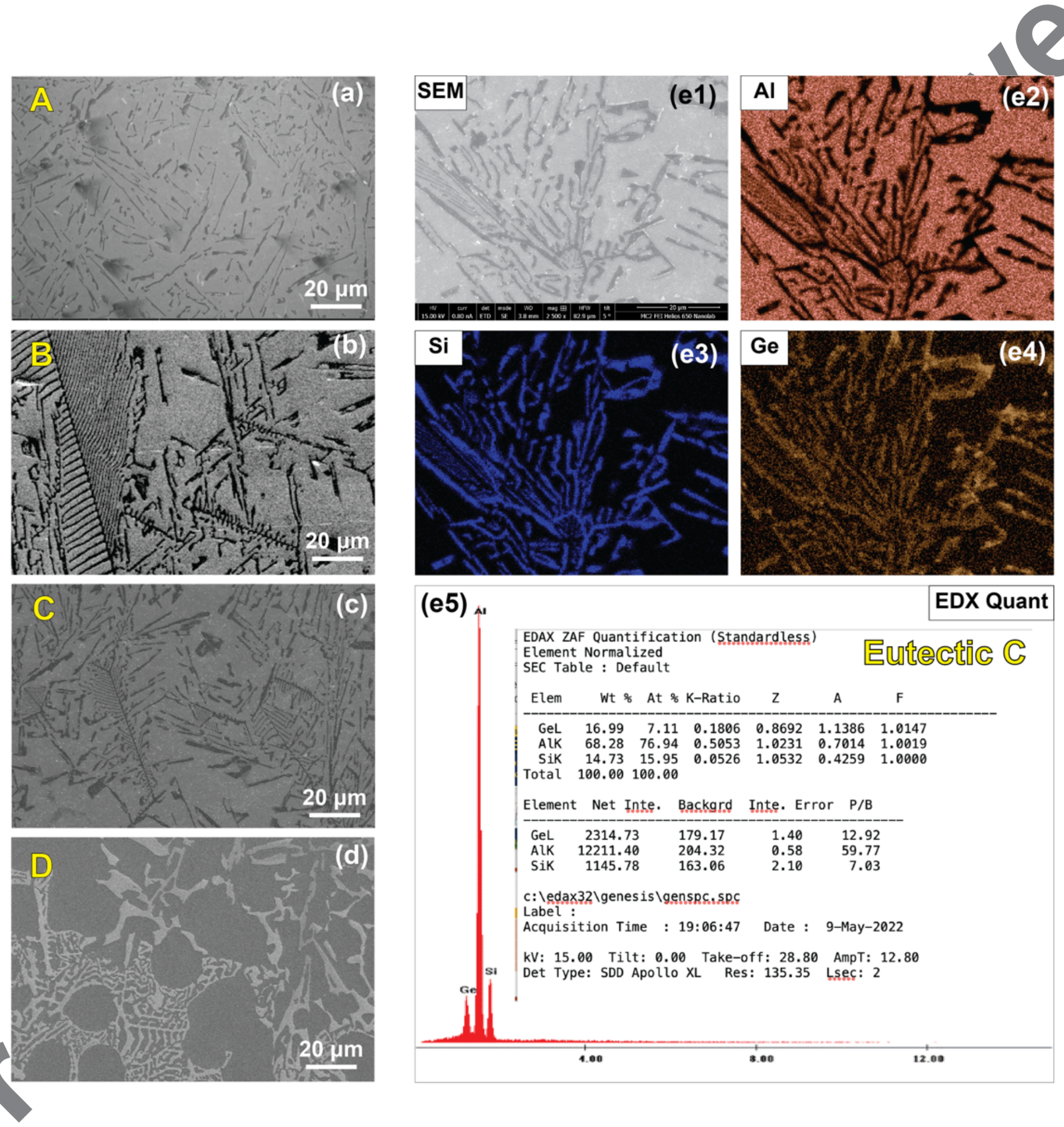

Figure S1: (a–d) SEM images showing arc melted microstructures of eutectics A, B, C, and D, that were subjected to laser rapid solidification. (e1–e4) SEM-EDX elemental maps confirmed the univariant eutectic [Al + Si–Ge] pathway for the ternary eutectics. Si and Ge are homogeneously distributed in the flake-like Si–Ge phases Here the maps from eutectic C have been displayed and (e5) elemental peaks as well as quantification data are provided.

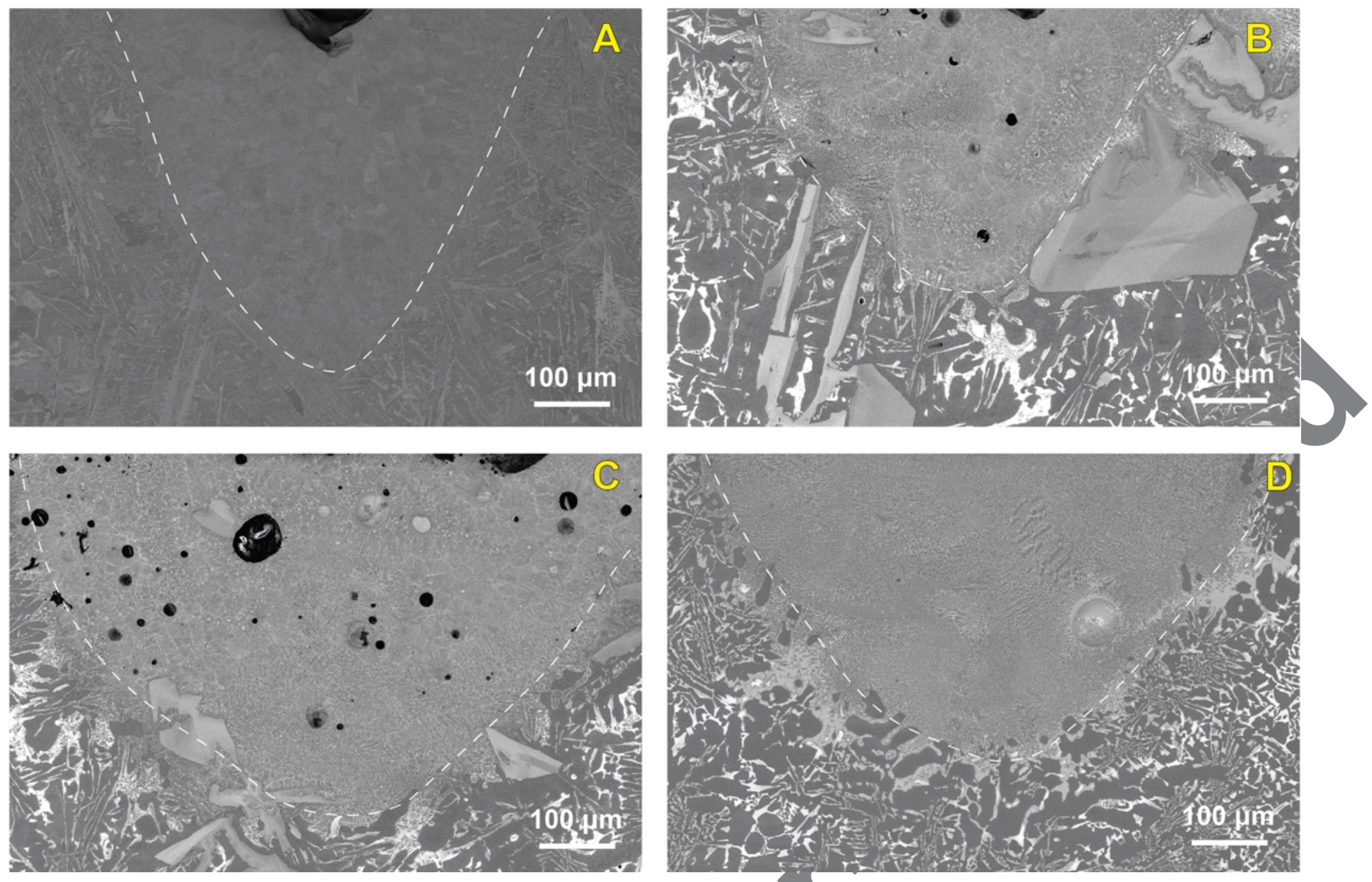

Figure S2: Representative melt pools obtained after laser surface remelting on the arc melted eutectics A–D.

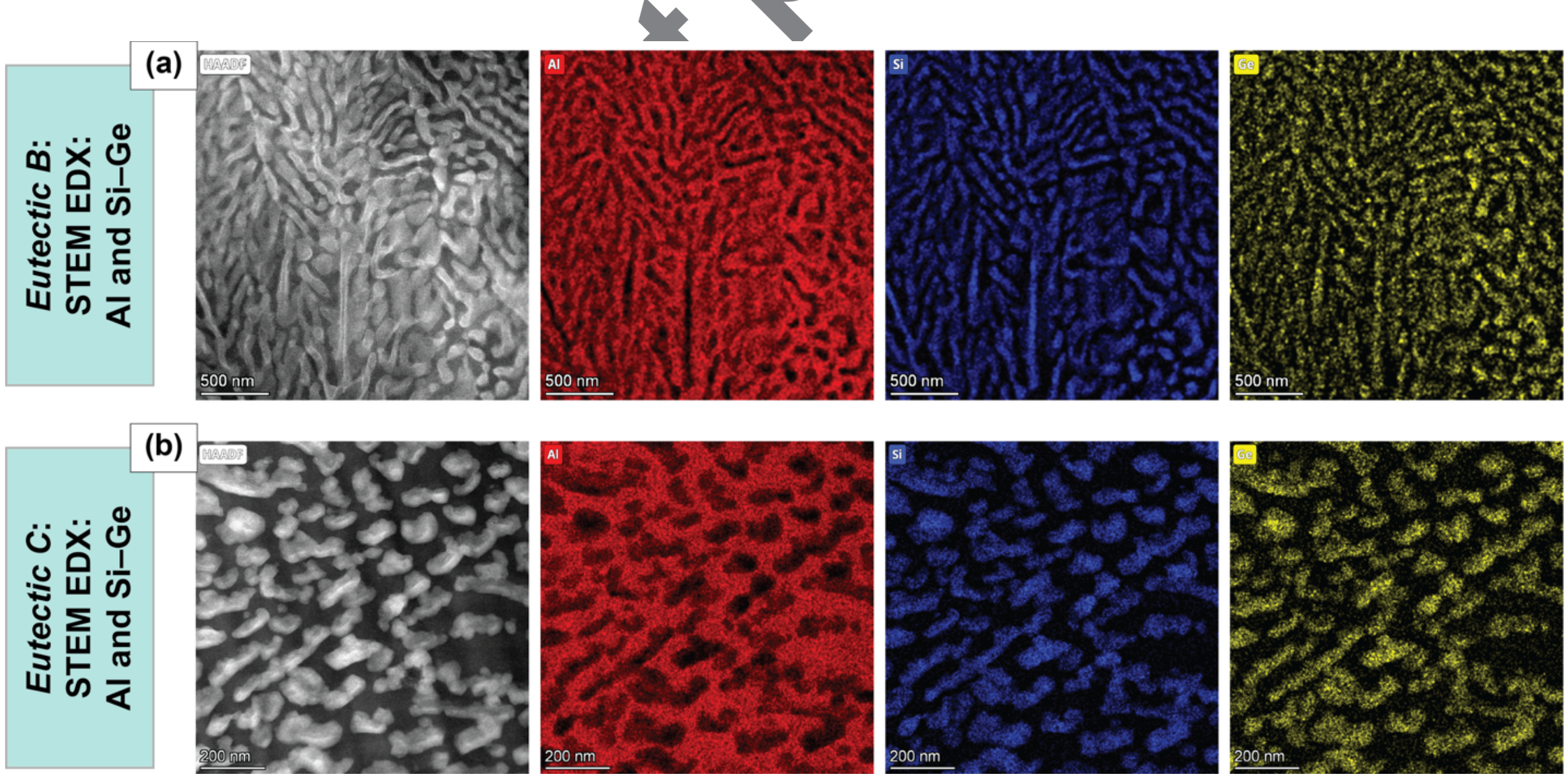

Figure S3: STEM-EDX elemental maps from ultrafine eutectics within melt pools from (a) eutectic B and (b) eutectic C. The maps confirm that the two-phase microstructure is prevalent across all the ternary eutectic compositions: Al-rich phase and Si–Ge-rich phase.

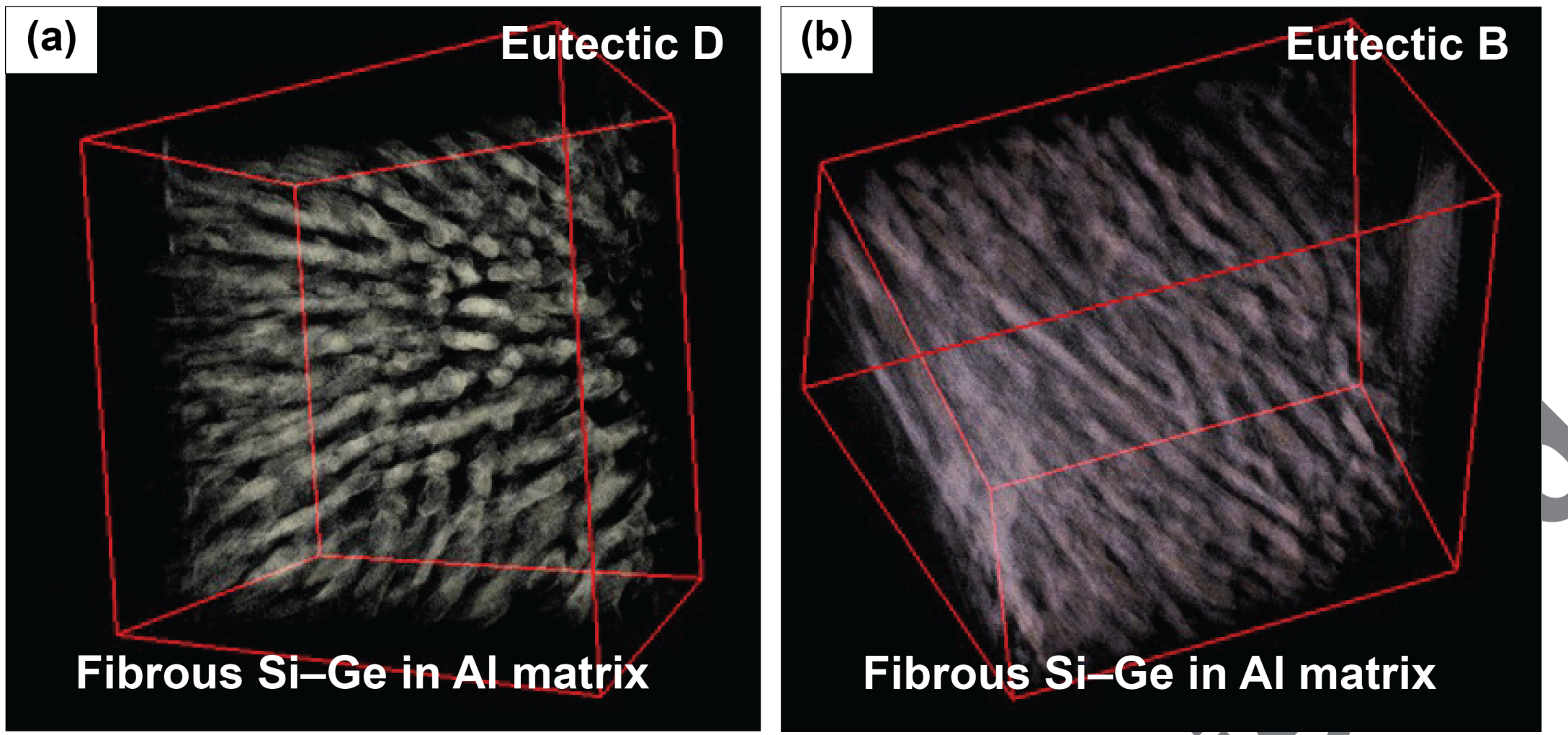

Figure S4: STEM-tomography-based three-dimensional reconstructions revealing the Si–Ge phases are distributed as fibrous/rod-like components in Al matix, which is consistent across all the ternary eutectics, as displayed for (a) eutectic D and (b) eutectic B, similar to binary Al–Si.

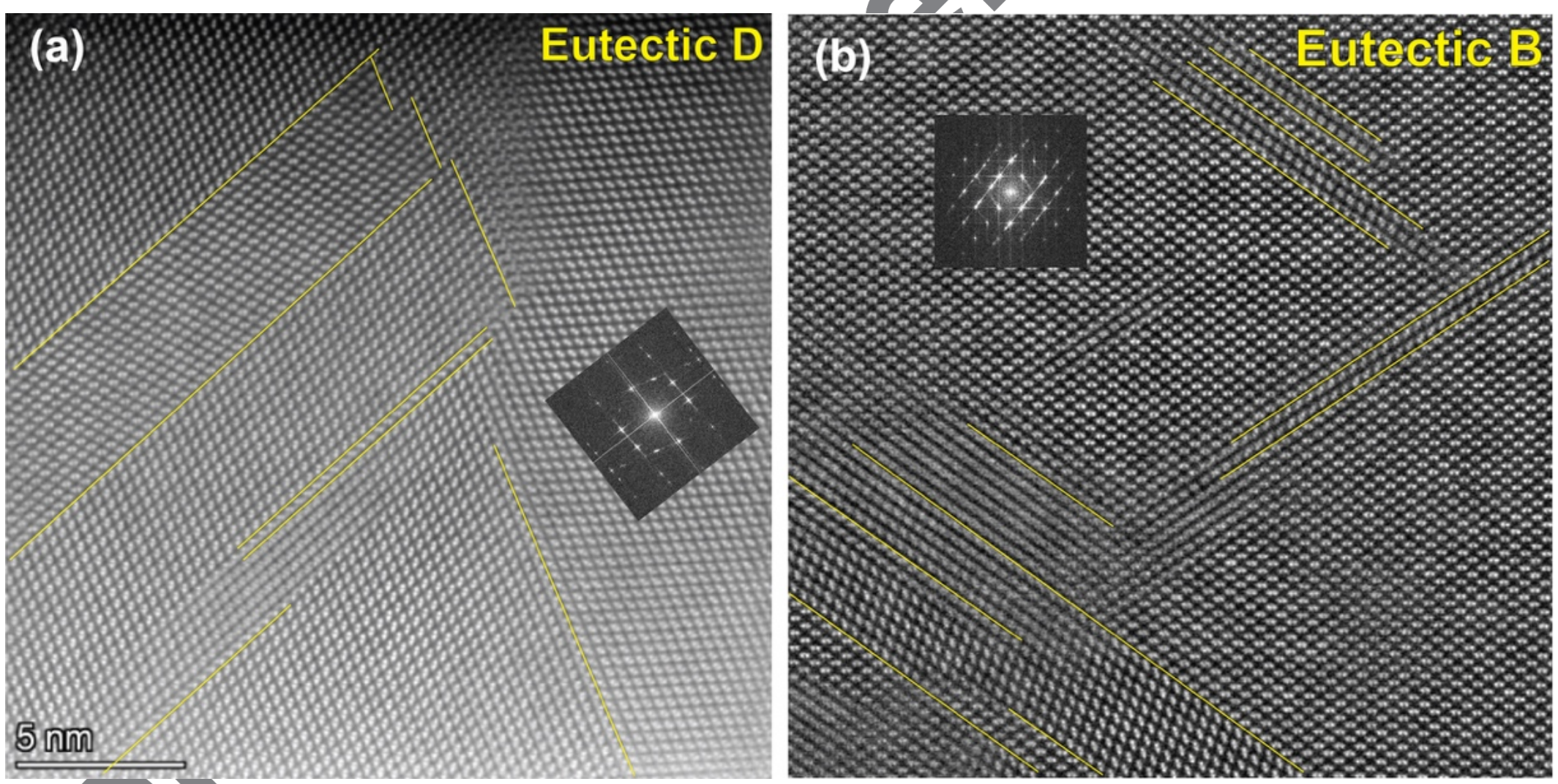

Figure S5: HAADF HR-STEM images reveal Si–Ge fibers with high density of growth twins across all the ternary eutectic compositions similar to that reported for Si fibers in binary Al–Si, here the images are displayed for (a) composition D and (b) composition B. Twin boundary spacing (i.e., twin thickness) and density increases with increasing Ge-content, i.e., from eutectics B to D.

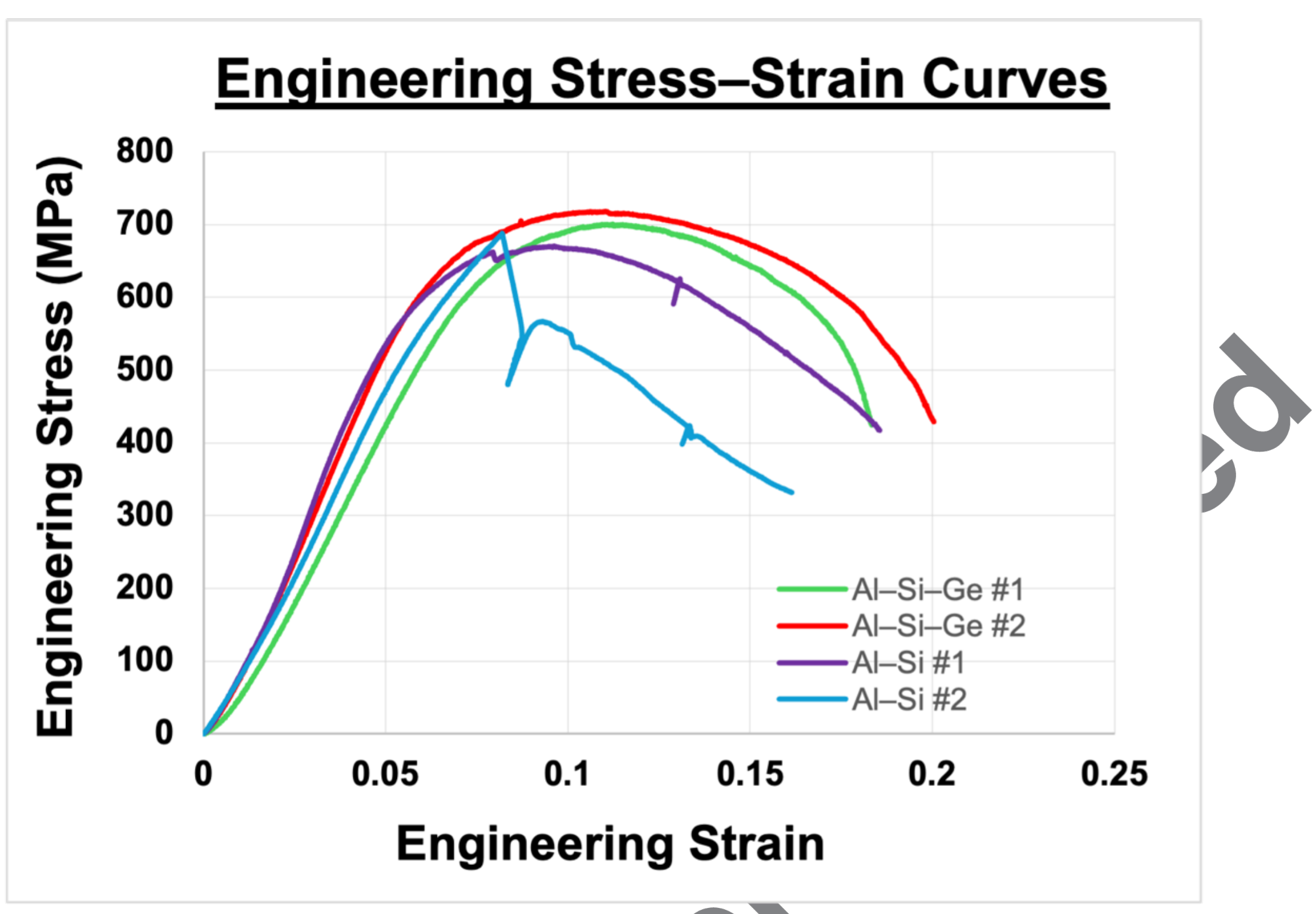

Figure S6: Engineering stress–strain curves from all the in situ SEM tensile tested samples (Al–Si–Ge eutectic D and Al–Si reference eutectic A) showing decent reproducibility. Certain load drops in the Al–Si curves can be attributed to premature cracking of Si fibers. It should be noted that the initial linear slope of the engineering curves should not be interpreted as the elastic modulus, because strain was inferred from gripper displacement rather than measured directly from the gauge section and therefore includes contributions from system compliance and local alignment effects. The key difference emerges after yielding – eutectic D consistently sustains a higher flow stress over a larger plastic strain range unlike eutectic A. Average UTS measured for eutectic D and eutectic A are measured as ~715 MPa and ~680 MPa, respectively. Notably, the Al–Si specimens exhibit discrete load drops during the early stages of plastic deformation. In situ SEM observations associate these events with the onset of localized Si-fiber cracking at approximately 2–3% plastic strain, indicating that damage accumulation begins substantially before final tensile failure.

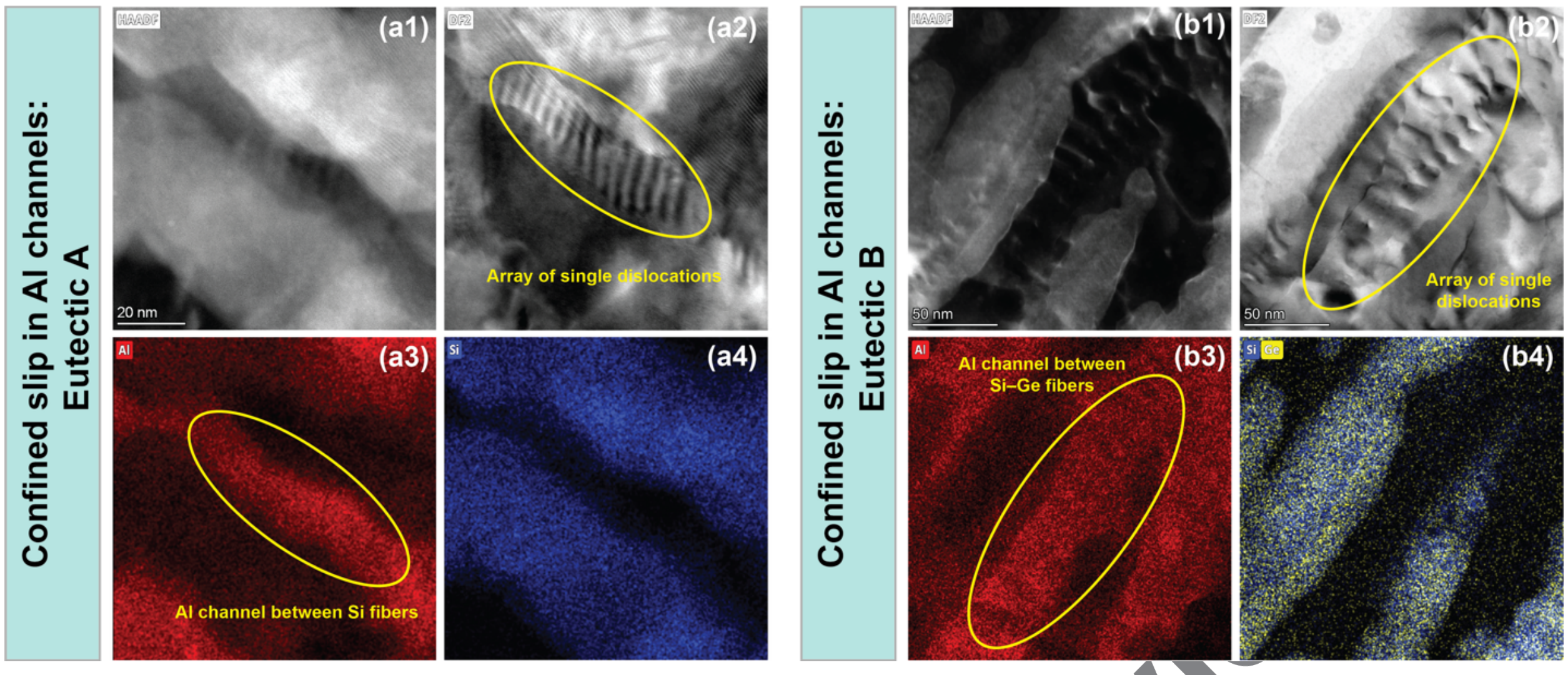

Figure S7: Array of single dislocations in Al channels embedded between Si- and Si–Ge-fibers revealed by BF-STEM and confirmed by HAADF-STEM and STEM-EDX maps for (a1–a4) eutectic A and (b1–b4) eutectic B. These are typical examples of confined slip reported in nano-scale heterostructures. In the main manuscript, we have reported a special type of confined slip for higher Ge-containing eutectics, which progressively takes place through low- and high-angle boundary formation, likely assisted by Si–Ge clusters in Al matrix.

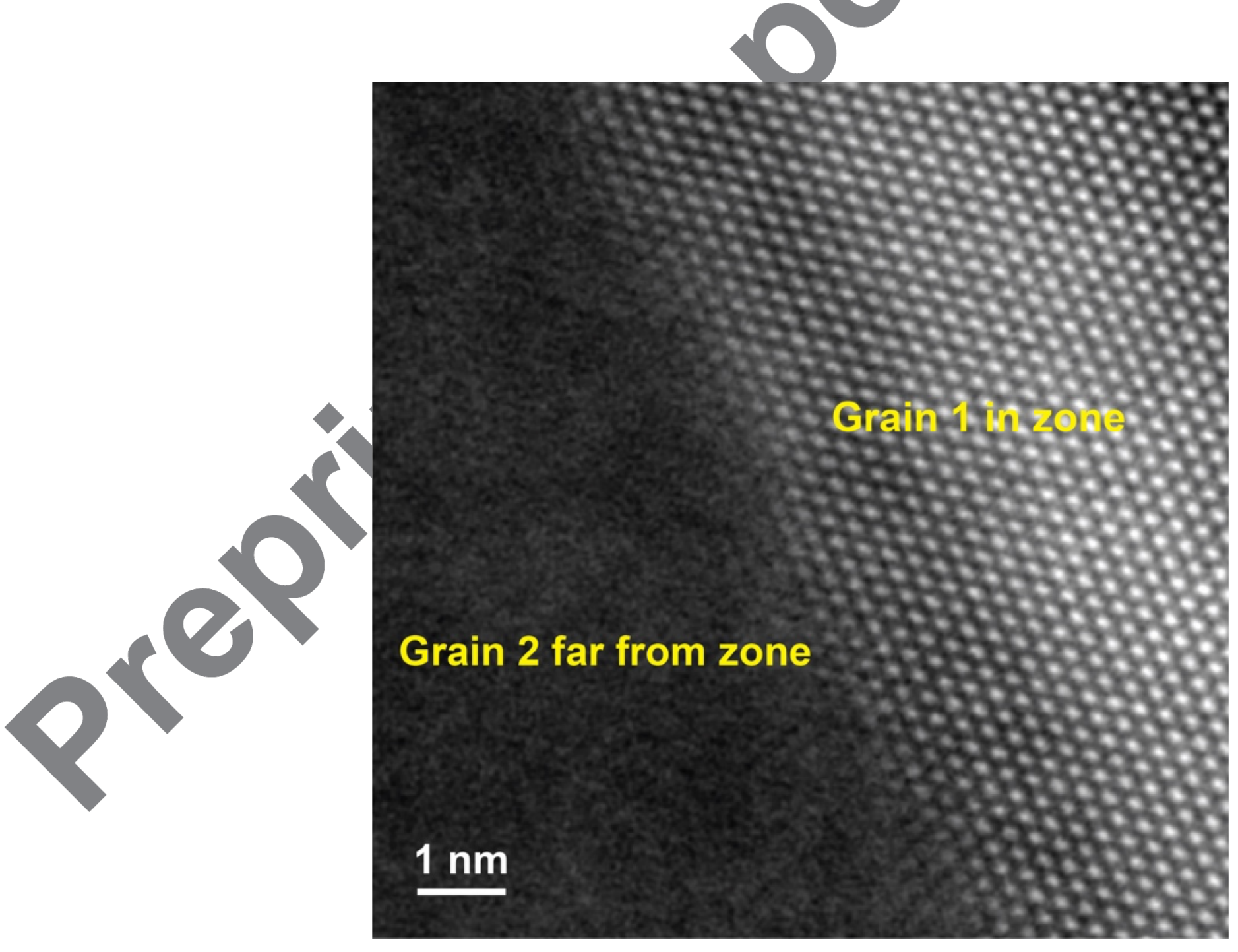

Figure S8: An example of high-angle boundary formation in Al-rich phase, where the two neighboring grains do not share common zone axis. As the grains are 3D, relative rotation of the grains often makes it difficult to index the high-angle boundaries properly using TEM.

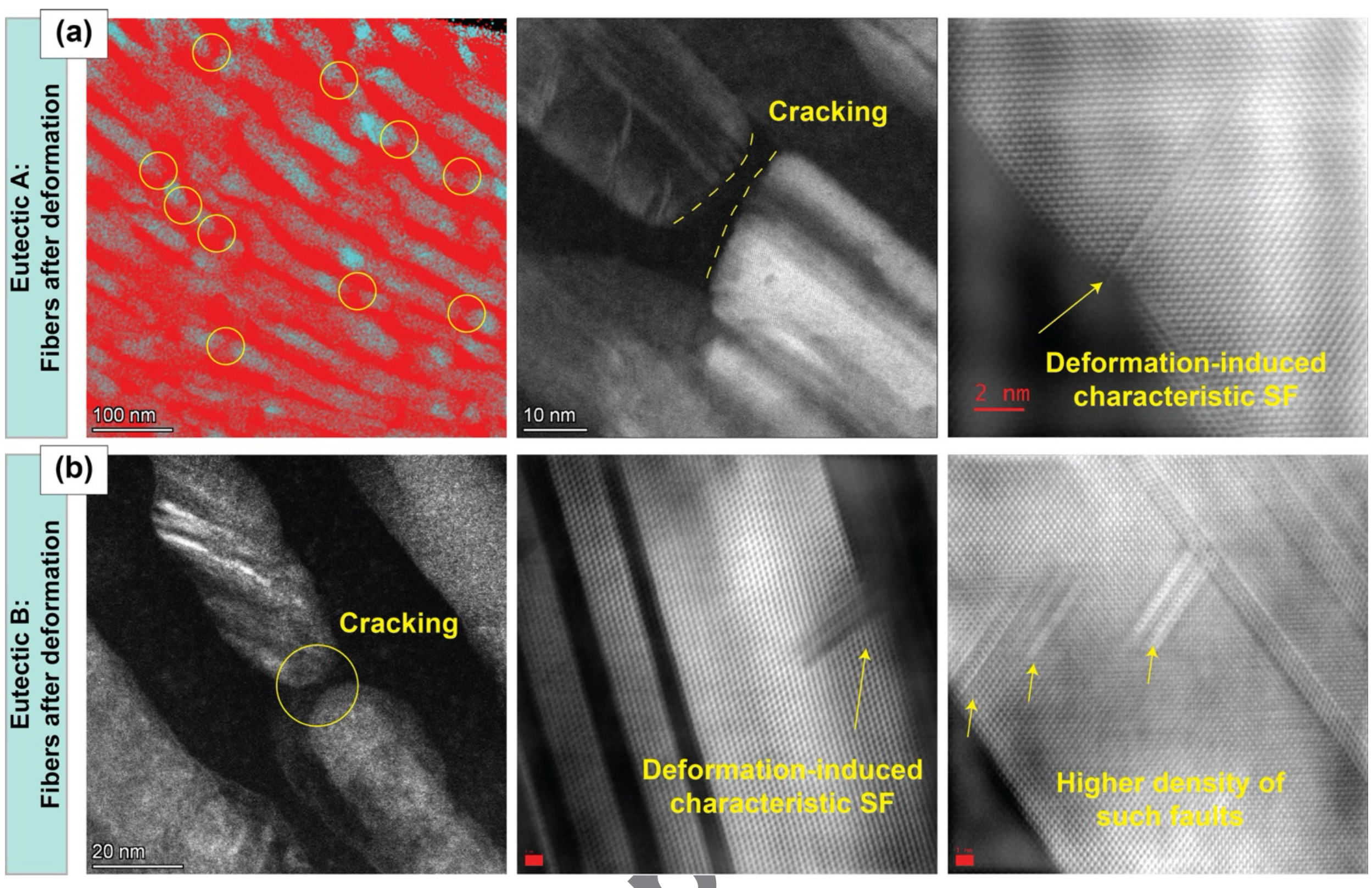

Figure S9: (a) Profound cracking in Si fibers after deformation in Al–Si reference eutectic A. At low magnification, we have used STEM-EDX map to identify the discontinuity in elongated Si fibers due to cracks since HAADF-STEM does not provide enough contrast to distinguish Al (Z=13) and Si (Z=14), while BF/DF STEM suffers from diffraction contrast. DF-STEM reveals such cracking at higher magnification. Although not as frequent as the Ge-containing eutectics, occasional presence of deformation-induced stacking faults (SFs) were recorded in Si fibers primarily at the interfaces at relatively less-strained regions. (b) In eutectic B, there is some observation of fiber cracking (not as profound as eutectic A). Presence of deformation-induced SFs becomes more prominent.

## Appendix A: Approximate finite-element analysis of strain heterogeneity across the compressed micropillar

A two-dimensional axisymmetric finite-element model was implemented using a custom Python code to estimate the spatial distribution of plastic deformation within the micropillar during compression. Numerical assembly and solution of the finite-element equations were performed using NumPy and SciPy, while post-processing was carried out using Pandas and Matplotlib. The tapered pillar was discretized using *108* four-node bilinear axisymmetric quadrilateral elements arranged as *6* elements across the radial direction and *18* elements along the pillar height. A $2 \times 2$ Gauss integration scheme was used within each element. The mesh was selected to capture the overall axial variation in plastic strain while maintaining computational efficiency. Because the objective of the calculation was to provide a semi-quantitative estimate of the local deformation level associated with different microscopy regions, rather than to resolve highly localized contact fields, no quantitative interpretation was made from individual-element maxima near the punch edge or fixed-base corner.

The experimentally measured geometry of eutectic D was reproduced, with an initial pillar height of $H_0 = 18.03\ \mu m$, top diameter of $d_t = 8.256\ \mu m$, and sidewall taper angle of $\alpha = 4°$. The local radius in the reference configuration was defined as

$$R(Z) = R_t + (H_0 - Z) \tan \alpha \quad \text{(A1)}$$

where $R_t = {}^{d_t}/_{2}$. This gives a calculated bottom diameter of approximately $10.78\ \mu m$. Axisymmetry was imposed along $R = 0$, such that

$$u_r(R = 0) = 0 \quad \text{(A2)}$$

The bottom surface was fully constrained in both the radial and axial directions,

$$u_r(Z = 0) = u_z(Z = 0) = 0, \quad \text{(A3)}$$

to approximate the mechanical continuity between the FIB-machined pillar and the underlying bulk specimen. The lateral pillar surface was traction-free.

Compression was imposed through the upper pillar surface using a prescribed axial displacement. For the experimentally applied nominal compressive strain of *30%*, the total displacement was

$$\Delta H = 0.30 H_0 = 5.409\ \mu m \quad \text{(A4)}$$

The corresponding macroscopic compressive true strain is

$$\varepsilon_{true} = -\ln\left(1 - \frac{\Delta H}{H_0}\right) \approx 0.357 \quad \text{(A5)}$$

The total displacement was applied incrementally in *45* loading steps to improve numerical convergence and to allow progressive evolution of the plastic deformation field.

Interaction between the flat diamond compression probe and the pillar top surface was represented through an effective tangential penalty resisting radial sliding of the top surface. The tangential penalty stiffness was defined as:

$$k_t = \mu \frac{E}{H_0} \tag{A6}$$

where $\mu = 0.1$ was used as the friction parameter. This treatment approximates the radial constraint associated with friction between the diamond probe and the pillar while retaining the computational efficiency of the axisymmetric model. It should be noted that this represents an effective frictional boundary condition rather than a full surface-to-surface Coulomb contact formulation.

The eutectic microstructure was treated as a homogenized isotropic elastoplastic continuum because the objective of the calculation was to determine the pillar-scale variation in deformation, rather than the local partitioning of strain between the Al-rich and Si–Ge-rich phases. The elastic response was described using

$$E = 75\ GPa, \upsilon = 0.3 \tag{A7}$$

Plastic yielding was described using isotropic $J_2$ plasticity with a von Mises yield criterion,

$$f(\sigma, \bar{\varepsilon}^p) = \sigma_{eq} - \sigma_y(\bar{\varepsilon}^p) \leq 0 \tag{A8}$$

where

$$\sigma_{eq} = \sqrt{\frac{3}{2} s : s} \tag{A9}$$

is the von Mises equivalent stress and *s* is the deviatoric stress tensor. The accumulated equivalent plastic strain was defined as:

$$d\bar{\varepsilon}^p = \sqrt{\frac{2}{3} d\varepsilon_p : d\varepsilon_p} \tag{A10}$$

The isotropic hardening response, $\sigma_y(\bar{\varepsilon}^p)$, was obtained from the experimentally measured true stress–true strain response of eutectic D [Figure 6b]. The elastic contribution to the measured true strain was removed according to

$$\varepsilon_p = \varepsilon_{true} - \frac{\sigma_{true}}{E} \tag{A11}$$

and the resulting true flow stress–true plastic strain data were represented by piecewise-linear interpolation. The input curve had an initial flow stress of approximately *560 MPa* and progressively increased toward approximately *780 MPa* at large plastic strain. Plastic correction at each integration point was performed using an incremental radial-return algorithm.

The nonlinear equilibrium equations were solved incrementally using a Newton–Raphson procedure. At each loading increment, the global residual force vector was expressed as:

$$\mathrm{R} = \mathrm{F_{ext}} - \mathrm{F_{int}} \tag{A12}$$

and displacement corrections were obtained from

$$K_t \Delta u = R \quad \text{(A13)}$$

where $K_t$ is the current tangent stiffness matrix. The geometry was updated after each converged load increment to account for the substantial change in pillar dimensions during compression. The constitutive update itself employed an incremental $J_2$ formulation.

Following completion of the *30%* compression step, the equivalent plastic strain was evaluated at the element integration points and averaged to the element centers. To facilitate direct comparison with microscopy locations identified on the undeformed pillar, the calculated fields were mapped using the reference coordinates (R,Z), with $Z = 0$ at the pillar base and $Z = 18.03\ \mu m$ at the initial top surface. The magnitude of the equivalent-plastic-strain gradient was subsequently evaluated as:

$$|\nabla \bar{\varepsilon}^p| = \sqrt{\left(\frac{\delta \bar{\varepsilon}^p}{\delta R}\right)^2 + \left(\frac{\delta \bar{\varepsilon}^p}{\delta Z}\right)^2} \quad \text{(A14)}$$

where the spatial derivatives were calculated numerically from neighboring element-center values in the reference configuration. The resulting strain-gradient magnitude therefore has units of $\mu m^{-1}$. For comparison with the TEM/STEM observations, equivalent plastic strain and strain-gradient values were spatially averaged over lower, middle, and upper regions of the pillar rather than using individual-element maxima. Values immediately adjacent to the probe edge and fixed-base corner were excluded from quantitative interpretation because these regions are particularly sensitive to contact idealization and mesh resolution.

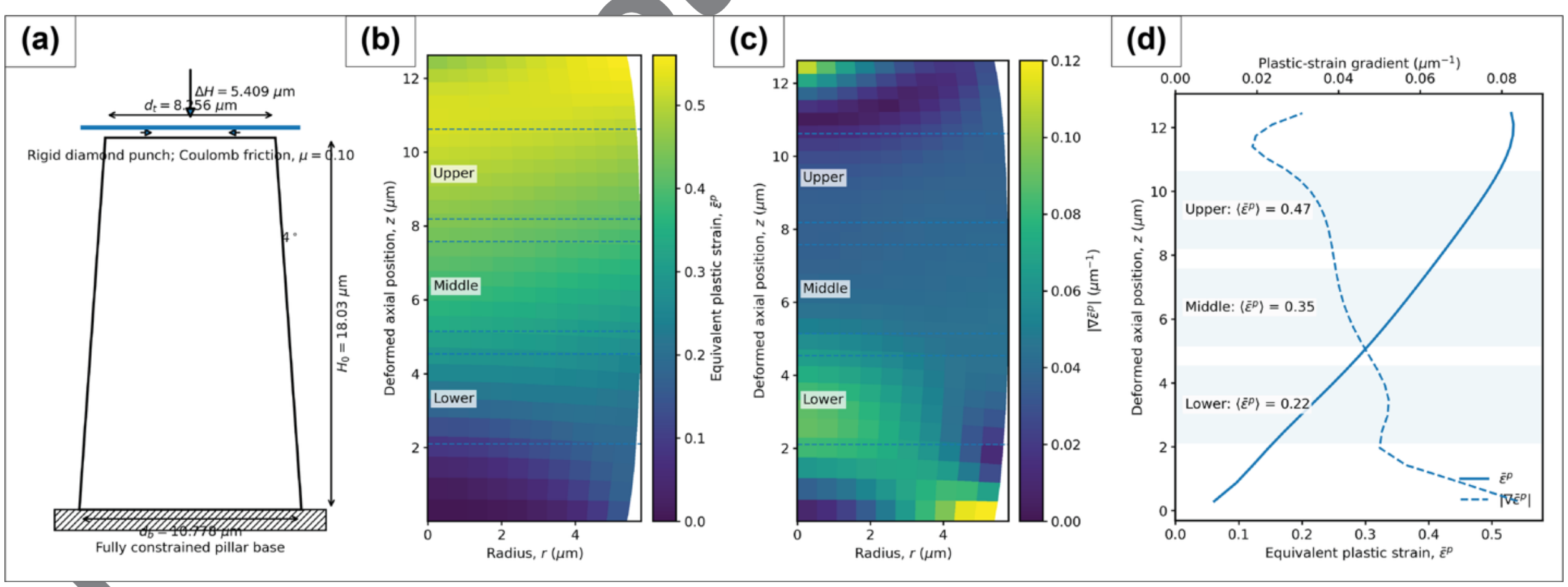

Figure A1: Finite-element estimation of strain heterogeneity during micropillar compression. (a) Axisymmetric tapered-pillar geometry and boundary conditions used in the simulation. (b) Equivalent plastic-strain distribution at *30%* nominal compression, showing the highest accumulated plastic strain in the upper pillar region and a progressive decrease toward the constrained base. (c) Corresponding plastic-strain-gradient magnitude, highlighting spatial variations associated with the pillar geometry and boundary constraints. Lower, middle, and upper regions indicate the locations used for microstructural correlation. (d) Axially averaged equivalent plastic strain and plastic-strain-gradient profiles, with average equivalent plastic strains of approximately *0.22*, *0.35*, and *0.47* in the lower, middle, and upper regions, respectively.

Although the present calculation is intentionally approximate and does not attempt to reproduce dislocation-level or phase-resolved deformation, it follows the general approach of using numerical modeling to estimate spatial variations in deformation within compressed micropillars [A1–A3] and to guide interpretation of site-specific postmortem microscopy.

## *Acknowledgements*

This work was funded by DOE, Office of Science, Office of Basic Energy Sciences with the grant number of DE-SC0016808. Arkajit Ghosh acknowledges the support of Rackham Predoctoral Fellowship by University of Michigan. Experimental characterization was performed in the Michigan Center for Materials Characterization ($(MC)^2$) at the University of Michigan-Ann Arbor. The authors acknowledge help of Dr. Tao Ma with the tomography experiment and thank Bruker scientists for allowing Arkajit Ghosh to use their Picoindenter at Minneapolis to perform pillar compression experiments. The authors are also grateful for the constructive discussions with Dr. Jian Wang from the University of Nebraska – Lincoln.

## *CRediT authorship contribution statement*

**Arkajit Ghosh:** Writing – original draft, Visualization, Methodology (testing and characterization), Investigation, Formal analysis, Validation, Software, Data curation, Conceptualization. **Amit Misra:** Writing – review & editing, Supervision, Project administration, Funding acquisition, Conceptualization.